\documentclass[useAMS,usenatbib]{mn2e}
\usepackage{natbib}
\usepackage{graphicx}
\usepackage{soul}
\usepackage{txfonts}
\usepackage[usenames,dvipsnames]{color}
\usepackage{supertabular}
\usepackage{longtable}
\bibpunct{(}{)}{;}{a}{}{,}

\title[Cepheid observations with the {\it MOST} satellite]
{  Observations of Cepheids with the {\boldmath $MOST$} satellite: Contrast between
Pulsation Modes  }
\author[N.~R. Evans, R. Szab\'o, A. Derekas et~al.]{N.~R. Evans$^{1}$, 
R. Szab\'o$^{2}$\thanks{E-mail: rszabo@konkoly.hu}, A. Derekas$^{2,3}$, 
L. Szabados$^{2}$, C. Cameron$^{4}$, J.~M. Matthews$^{5}$, \newauthor
D. Sasselov$^{1}$, R. Kuschnig$^{5,6}$, J.~F. Rowe$^{7}$, D.~B. Guenther$^{8}$,
A.~F.~J. Moffat$^{9}$, \newauthor S.~M. Rucinski$^{10}$, W.~W. Weiss$^{6}$\\
$^{1}$Harvard-Smithsonian Astrophysical Observatory, MS 4, 60 Garden St., Cambridge, MA 02138, USA\\
$^{2}$Konkoly Observatory, Research Center for Astronomy \& Earth Sciences,
Konkoly Thege Mikl\'os \'ut 15-17, H-1121 Budapest, Hungary\\
$^{3}$ELTE Gothard Astrophysical Observatory, H-9704 Szombathely, Szent Imre herceg \'ut 112, Hungary\\
$^{4}$Department of Mathematics, Physics \& Geology, Cape Breton University, 1250 Grand Lake Road,
Sydney, Nova Scotia, Canada, B1P 6L2 \\
$^{5}$Department of Physics and Astronomy, University of British Columbia, Vancouver, BC V6T1Z1, Canada\\
$^{6}$University of Vienna, Institute for Astronomy, T\"urkenschanzstrasse 17, A-1180 Vienna, Austria\\
$^{7}$NASA Ames Research Center, Moffett Field, CA 94035, USA\\
$^{8}$Department of Astronomy and Physics, St. Mary's University, Halifax, NS B3H 3C3, Canada\\
$^{9}$D\'epartment de physique, Universit\'e de Montr\'eal C.P. 6128, Succursale Centre-Ville, Montr\'eal, QC H3C 3J7, Canada\\
$^{10}$Dept. of Astronomy and Astrophysics, University of Toronto, 50 St George Street, Toronto, ON M5S 3H4, Canada}

\begin{document}

\date{Accepted;  Received; in original form }
\maketitle
\label{firstpage}

\begin{abstract}The quantity and quality of satellite photometric data strings is
  revealing details in Cepheid variation at very low levels.
  Specifically, we observed a Cepheid pulsating in the fundamental
  mode  and one pulsating in the first overtone with the Canadian 
  {\it MOST} satellite. The 3.7-d period fundamental mode pulsator (RT~Aur) has 
  a light curve that repeats precisely, and can be modeled by a Fourier 
  series very accurately. The overtone pulsator (SZ~Tau, 3.1 d 
period) on the other 
  hand shows light curve variation from cycle to cycle which we characterize 
  by the variations in the Fourier parameters. We present arguments that we are 
  seeing instability in the pulsation cycle of the overtone pulsator, and
  that this is also a characteristic of the $O-C$ curves of overtone pulsators.
On the other hand, deviations from cycle to cycle as a function of
pulsation phase follow a similar
pattern in both stars, increasing after minimum radius.   In summary,
pulsation in the overtone pulsator is less stable than that of the
fundamental mode pulsator at both long and short timescales.
\end{abstract}


\section{Introduction}

Classical Cepheid variable stars have light and velocity variations
which repeat very precisely.  This is in contrast to variations from
cycle to cycle sometimes seen in RR~Lyrae stars (Szab\'o et al. 2010). 
and Type~II Cepheids (Sterken \& Jaschek 1996).
In fact it is the  close  repeatability in classical Cepheids which
allows us to watch them evolve, that is to change their periods as
they move through the instability strip in the HR diagram.  

We  have a few examples of changes in the light curves of Cepheids, 
partly due to new precision and long duration of observations. V473~Lyr 
\citep{betal86} has a large change in amplitude over a period of 
about 4 years, resembling the Blazhko effect (Moln\'ar and Szabados 2014). 
Polaris itself has a variable amplitude which now begins 
to look cyclic, i.e. pulsation not evolution related (Arellano Ferro 
1983; Bruntt et al. 2008). Extensive observations from the {\it WIRE} 
satellite have helped to establish this behavior in Polaris. 

The field of view of the {\it Kepler} satellite contains numerous 
RR~Lyr stars but only one classical Cepheid, V1154~Cyg. However the
extensive photometry of this star has produced an intriguing result
\citep{der12}.  Typical behavior of the period variations is a change in
the period of 
approximately 20 minutes which lasts for approximately 15
cycles, but then a return to the mean period.   
Results for classical Cepheids from the {\it CoRoT} satellite by Poretti 
are expected shortly.  

Szabados (1983) has discussed period changes in Cepheids, both
fundamental mode pulsators and overtone pulsators (low amplitude `s'
Cepheids).  Based on his data, Evans, Sasselov \& Short (2002) have
suggested that the rapid period change in Polaris (an overtone
pulsator) is a characteristic of this group, a larger period
instability than found in fundamental mode stars.  

In order to follow up the discovery of a low level of period
instability in the extensive {\it Kepler} data in V1154~Cyg and to
further investigate differences in period instability between
fundamental and overtone mode pulsators, we have obtained sequences 
of photometry with the {\it MOST} ({\it M}icrovariability and 
{\it O}scillations of {\it St}ars) satellite. The target stars were 
RT~Aur (fundamental mode, $P = 3.7$~d, $\langle V \rangle = 5.5$ mag ,  
$V$ amplitude of 0.80 mag \footnote{http://www.astro.utoronto.ca/DDO/research/cepheids/}, 
F8 Ib \footnote{http://simbad.u-strasbg.fr/simbad/}) 
and SZ~Tau (overtone mode, $P = 3.1$~d, $\langle V \rangle = 6.4$ mag,  $V$ amplitude of 0.33 mag
F7 Ib). 
For comparison, V1154~Cyg is a fundamental mode pulsator with $P = 4.9$~d
and amplitude in $V$ of 0.40 mag.


\section{{\boldmath $MOST$} observations}\label{obs}

The {\it MOST} satellite is a photometric satellite fully described in 
Walker et~al. (2003), with the first science presented in Matthews et~al. 
(2004).  

For SZ~Tau, observations were obtained in November 2012 
(JD\,2\,456\,238-JD\,2\,456\,257) resulting in a continuous data set
covering 19 days with a cadence of 1 minute. 
For RT~Aur, the observations were carried out in December 2012
(JD\,2\,456\,278-JD\,2\,456\,300). For this target, the 22~day-long observations 
were interleaved with another target, resulting in gaps in the light curve.  

Data reduction was done following the standard steps outlined in Rowe
et al. (2006a, 2006b). 
Data can contain an artifact due to scattered earthshine in the {\it MOST}
101 minute orbit, resulting in a signal with multiple peaks near 14~c/d. 
The SZ~Tau data were reprocessed beyond the standard processing to greatly 
reduce this signal. In RT~Aur, the effect was not as prominent, but
ultimately they were reprocessed also.  Along with RT~Aur a `comparison 
star', HD\thinspace45237 was observed. This in principle would allow us to 
assess the quality of {\it MOST} photometry by comparing a relatively quiet 
star's light variations to that of our Cepheids. 
HD\thinspace45237, which is a K0~IV 7th magnitude star, shows minor variations, 
totaling less than 0.01 magnitude. Specifically, it shows a low-amplitude, 
`Gaussian-shape' brightening event in the middle of our observing session 
lasting for 6 days. Other than that, the frequency spectrum of the star nicely 
shows the alias frequencies due to the sampling around 14~c/d. We did not 
 use  this star further in our analysis.  


\begin{figure*}
\includegraphics[angle=270,width=8cm]{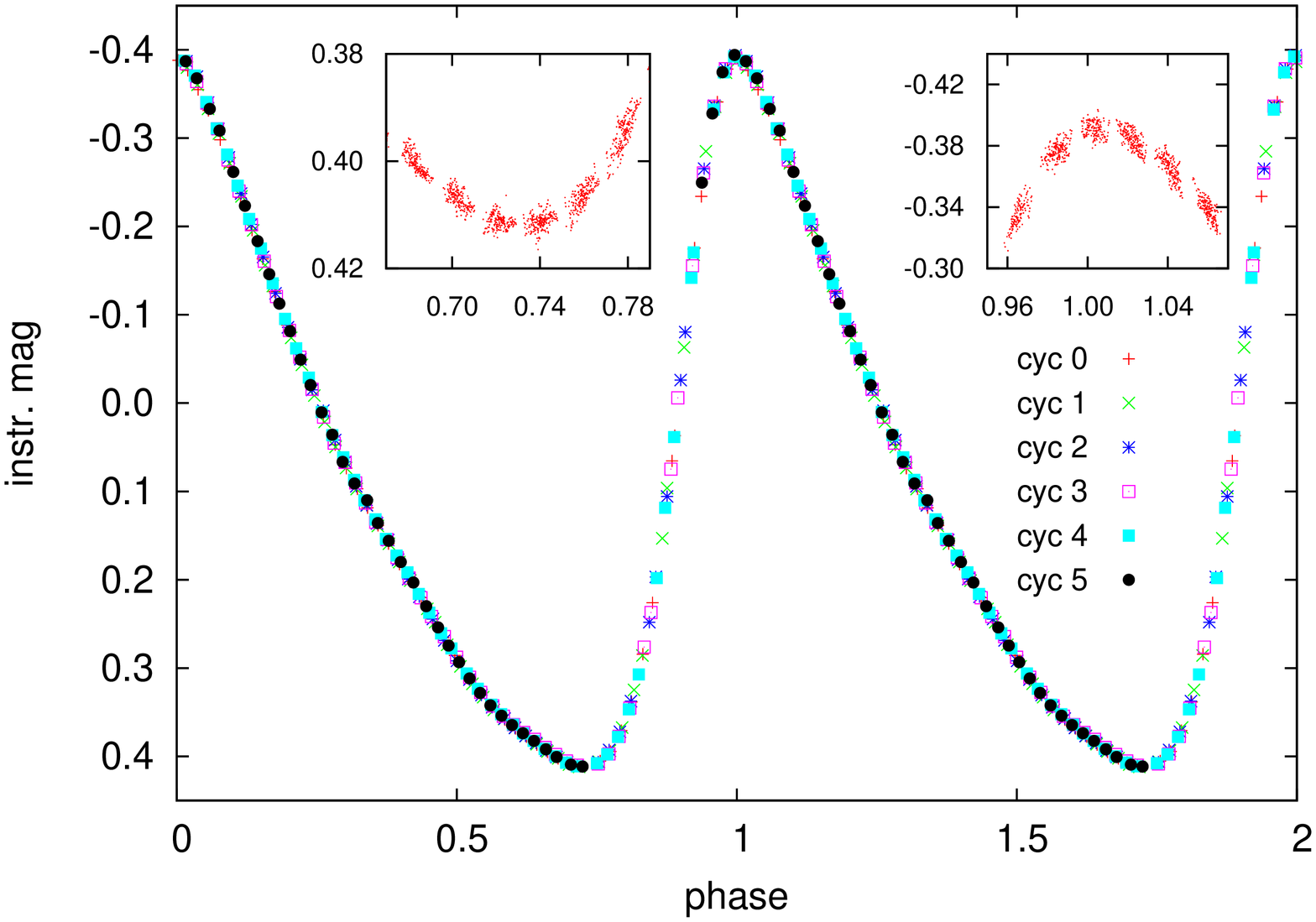}\includegraphics[angle=270,width=8cm]{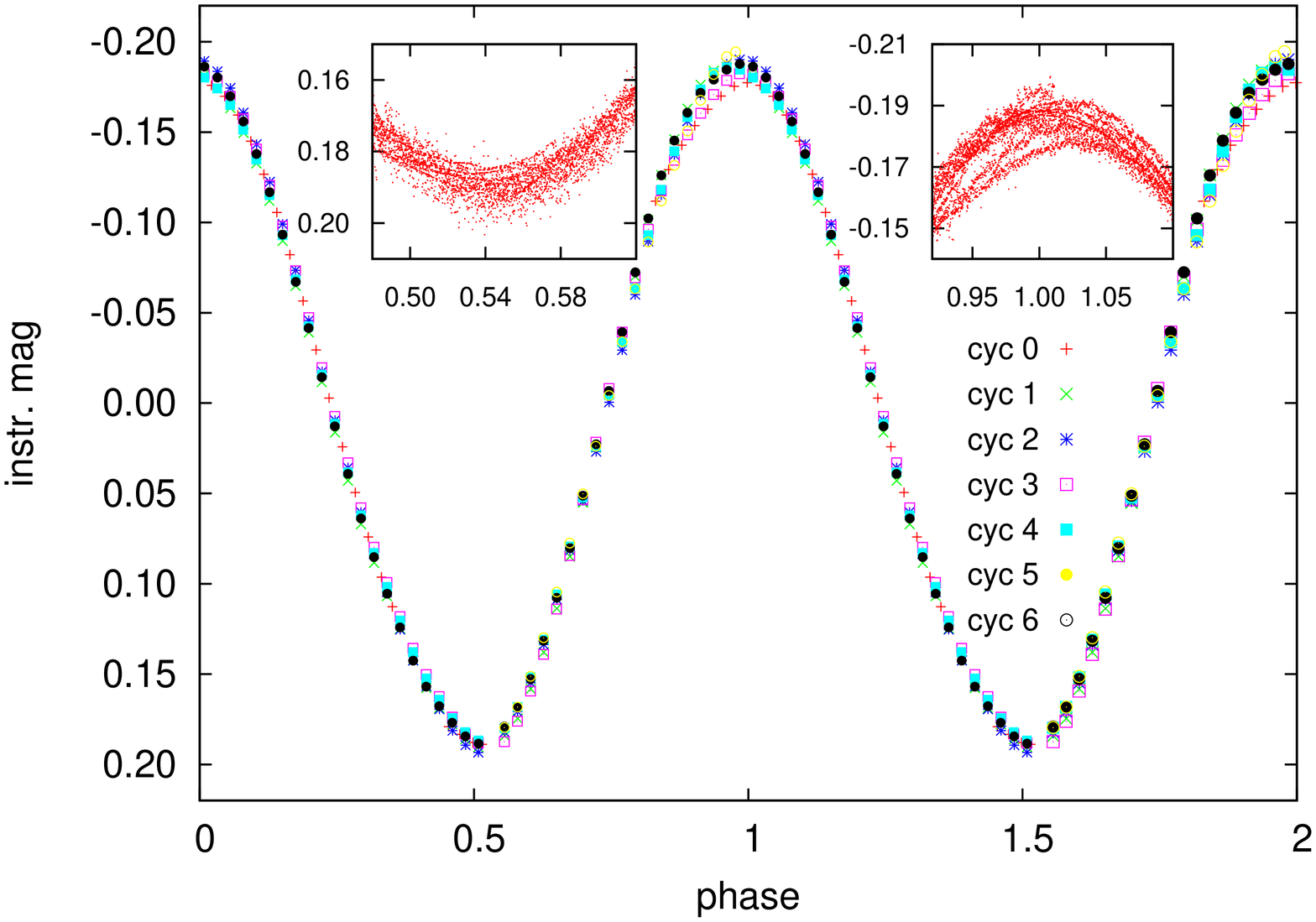}
\caption{{\it MOST} light curves of two Cepheids phased by pulsation period: 
the fundamental mode pulsator RT~Aur (left panel) and SZ~Tau (right panel) 
which pulsates in the first overtone. Note the different scale on the y axis. The insets show 
blow-ups of the maxima and minima indicated by the rectangles.  The data have been binned with a 0.075$^d$ bin size. A comparison of the two 
panels clearly shows that the pulsational cycles of RT~Aur 
repeat regularly, while there is a substantial deviation in the light 
variation of SZ~Tau, especially 
around and before the pulsational maximum. 
The full color plot is included in the electronic version.}
\label{mostc}
\end{figure*}


\begin{figure*}
\includegraphics[angle=270,width=5.5cm]{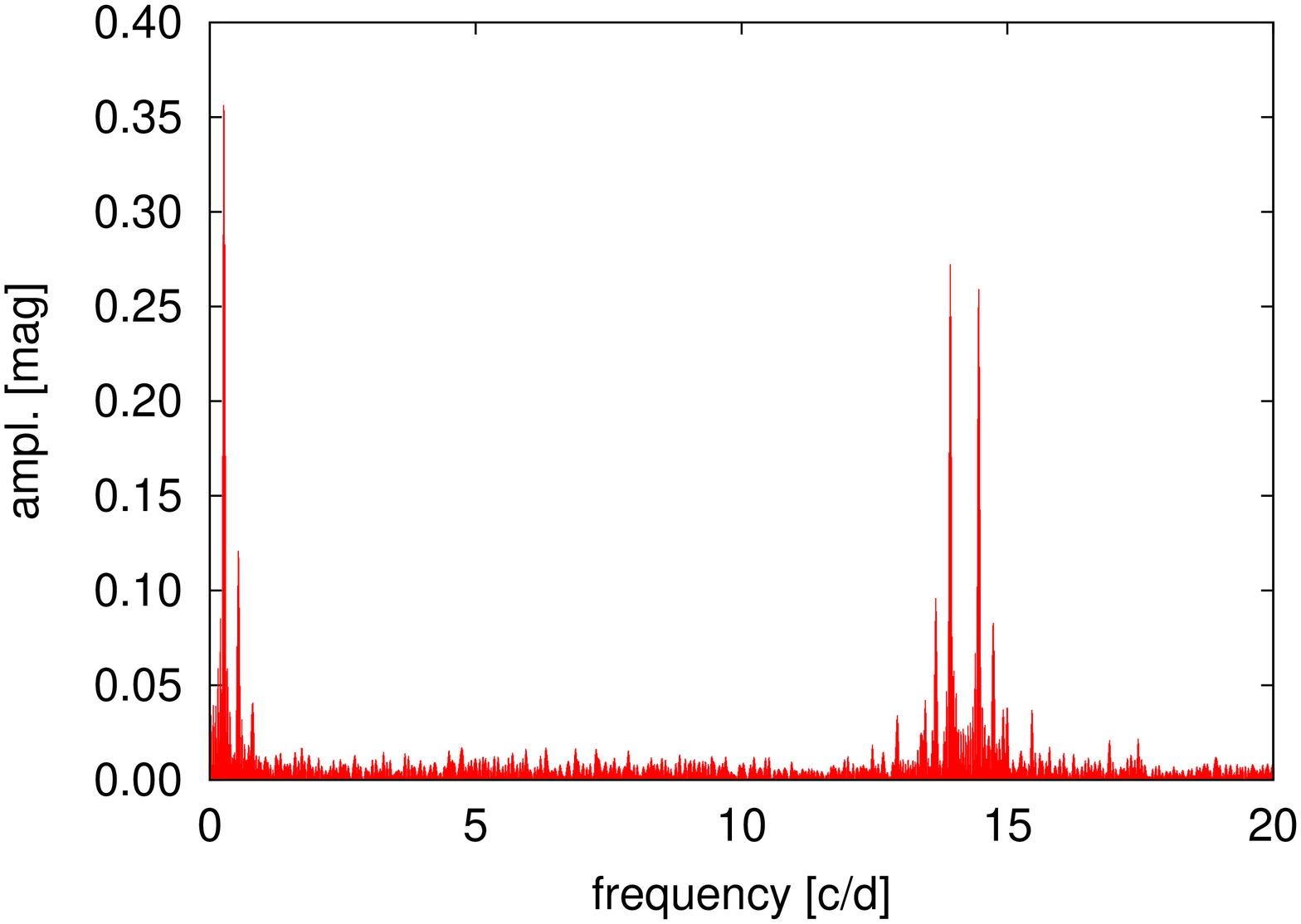}
\includegraphics[angle=270,width=5.5cm]{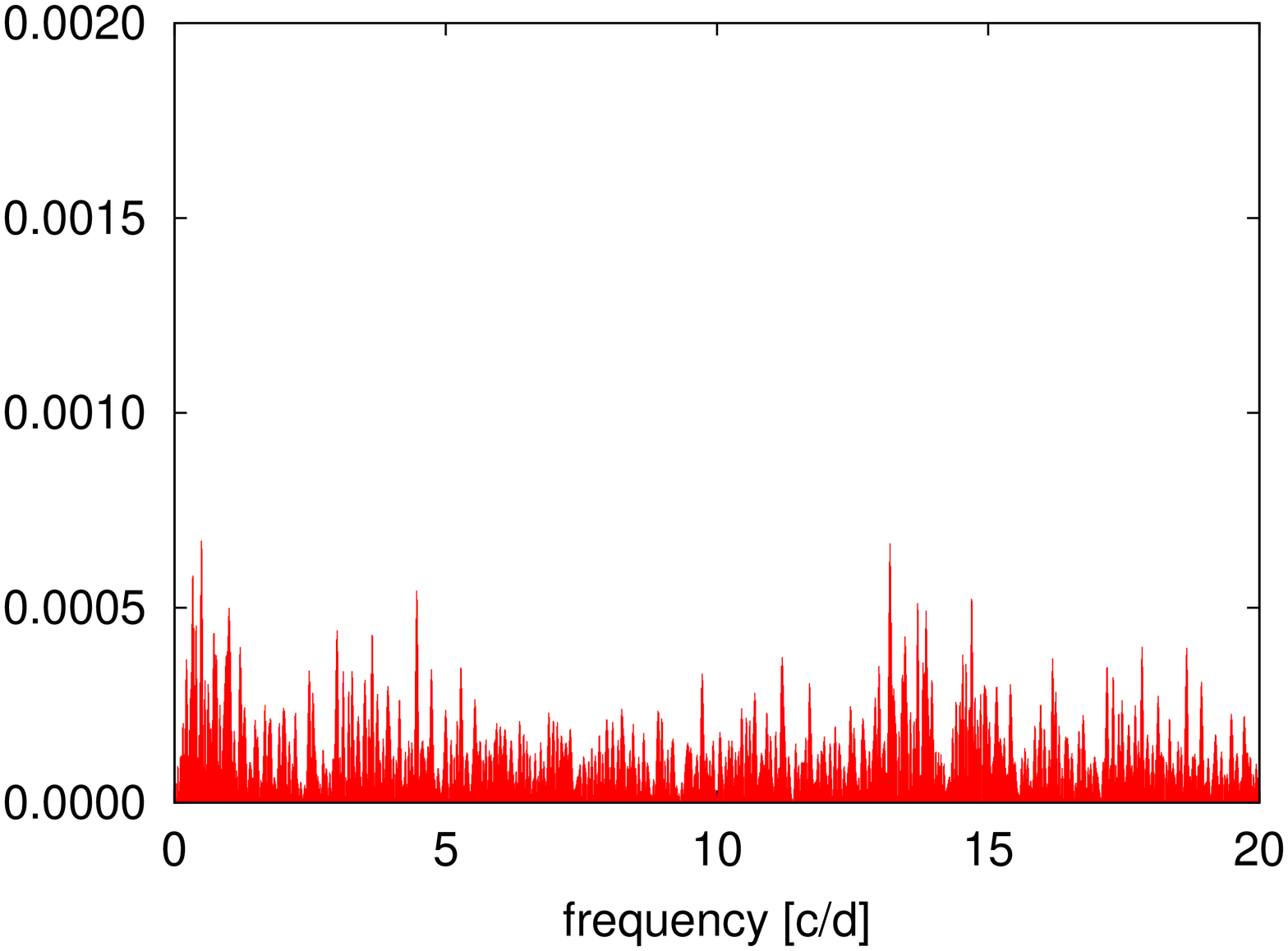}
\includegraphics[angle=270,width=5.5cm]{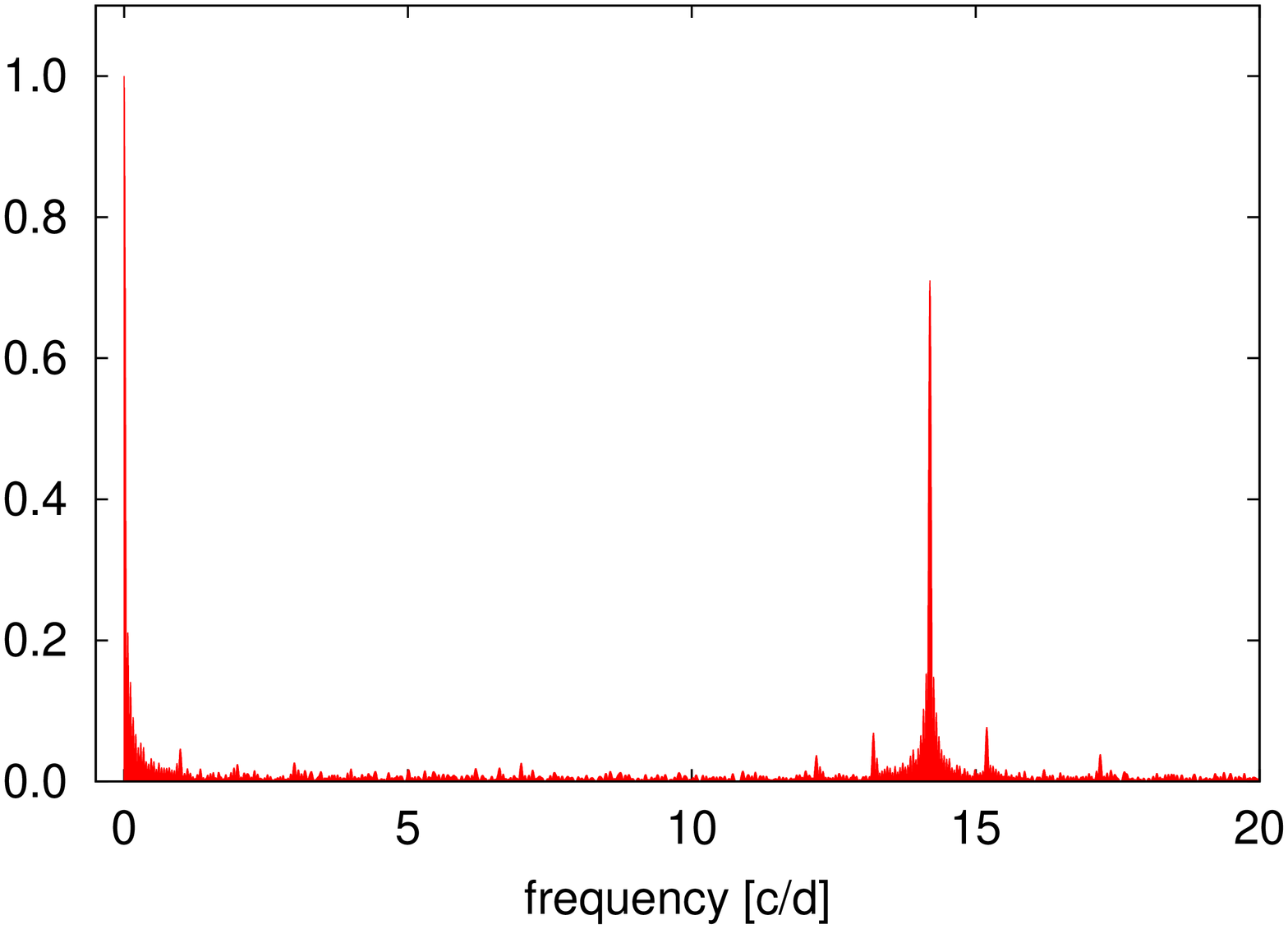}
\caption{Fourier spectrum of RT~Aur (left panel); the same, but 
prewhitened by the 10 frequencies in Table~\ref{rtfreqs} (the pulsational
frequency and its harmonics) (middle panel); spectral window of the 
RT~Aur observations of {\it MOST} (right panel). The feature 
 near 14 c/d is due to  interleaving the observations with another 
target.}
\label{rtFourier}
\end{figure*}

\begin{figure*}
\includegraphics[angle=270,width=5.5cm]{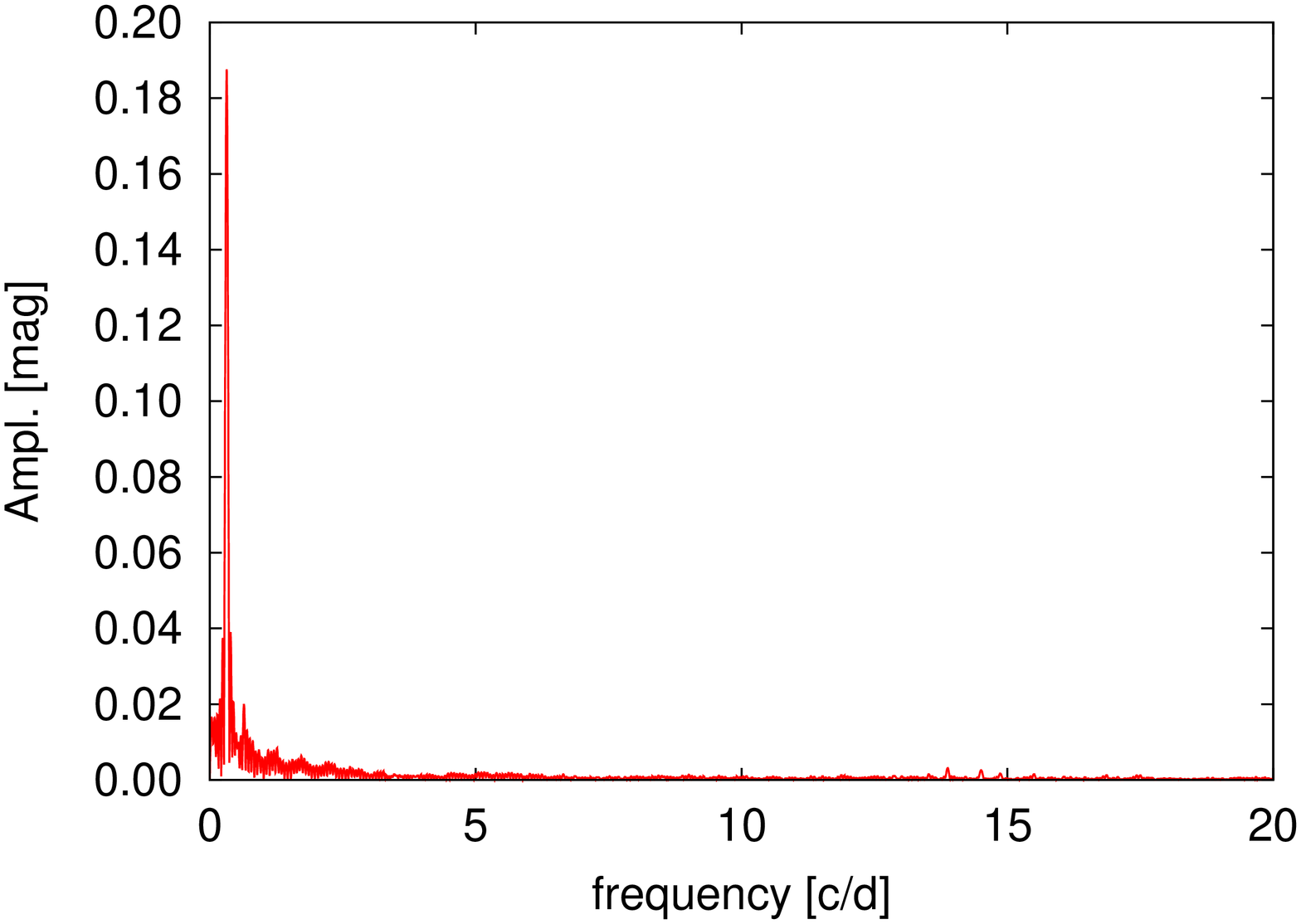}
\includegraphics[angle=270,width=5.5cm]{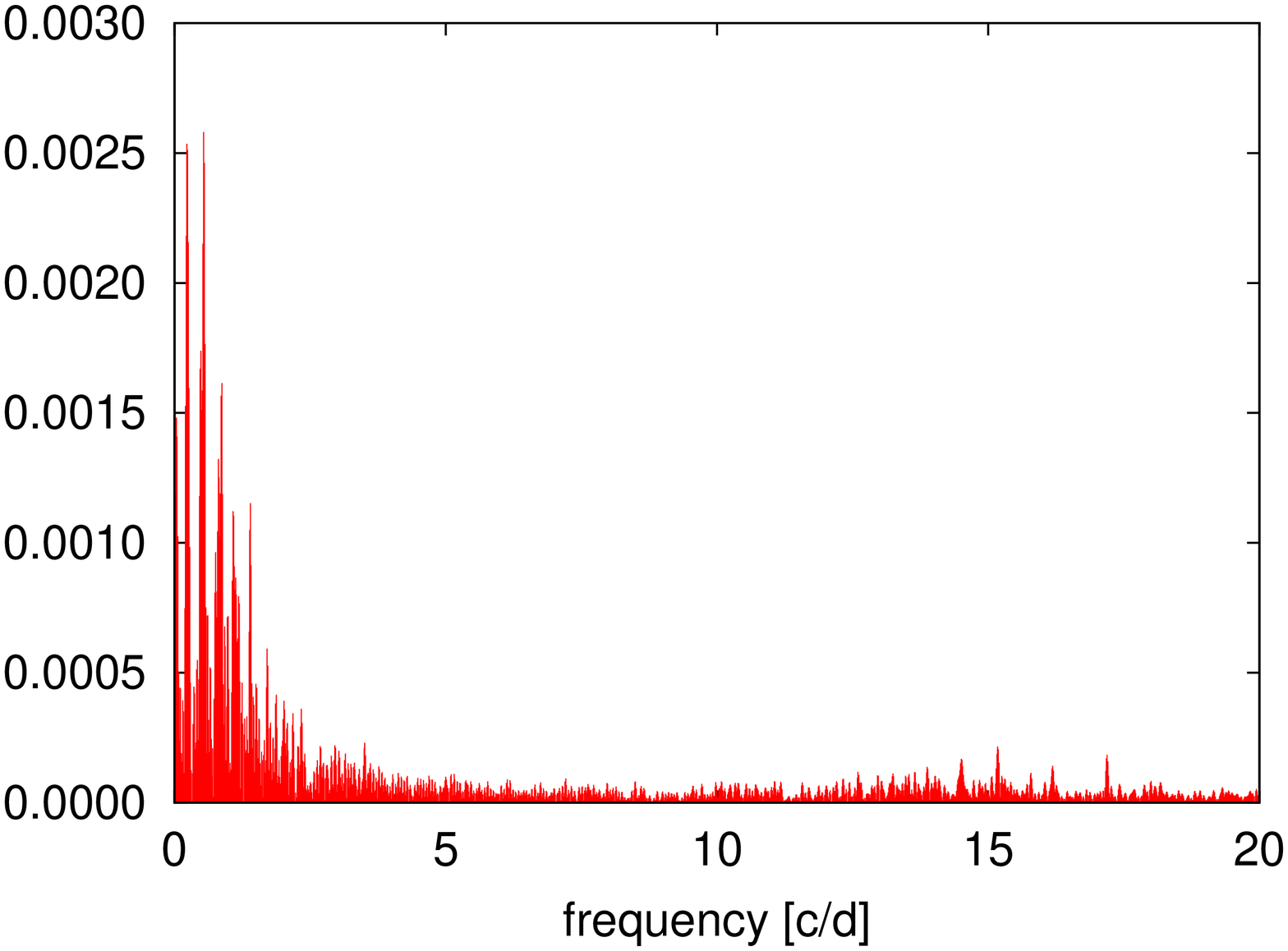}
\includegraphics[angle=270,width=5.5cm]{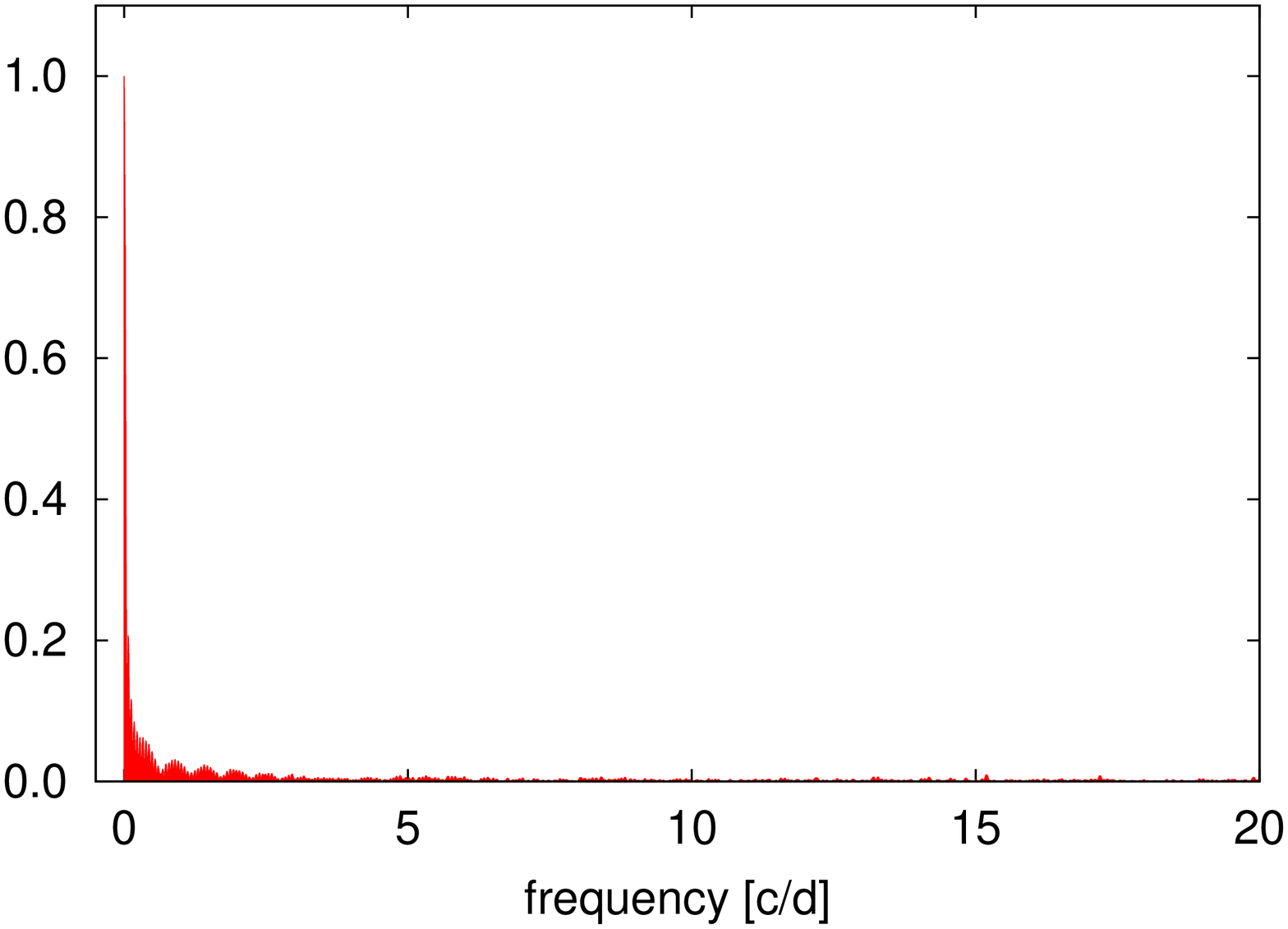}
\caption{Fourier spectrum of SZ~Tau (left panel); the same, but 
prewhitened by the 9 frequencies in Table~\ref{szfreqs} (the pulsational
frequency and its harmonics) (middle panel); spectral window of 
the SZ~Tau observations of {\it MOST} (right panel).}
\label{szFourier}
\end{figure*}

\section{Light curve analysis}

A difference between the light curves of RT~Aur and SZ~Tau is
immediately obvious from the first plots of the light curves.
Figure~\ref{mostc} shows the data of RT~Aur phased by a period of 3.7348~d
and SZ~Tau phased by a period of 3.149407~d, respectively.
The RT~Aur light curve repeats very precisely, 
notably at maximum light.  On the other hand, 
SZ~Tau shows tight sequences at maximum for each cycle  which, however, vary in brightness
from cycle to cycle. Similar but less pronounced behavior is seen 
at minimum light.  



The next step is to investigate the departures of the data from a
strict cycle-to-cycle repetition. In Fig.~\ref{mostc} we plot the 
consecutive pulsational cycles with different symbols and colors.
This figure shows that 
while the pulsational cycles of RT~Aur repeat regularly, there is a 
considerable deviation in the light variation of SZ~Tau, especially 
around and before the brightness maximum.

The rest of this discussion will be to quantify the contrasting
behavior between the fundamental and the overtone pulsators.

The first step in the analysis is to fit a Fourier 
polynomial to the data:

\begin{equation}
\label{foupol}
m = A_0 +  \displaystyle\sum_{i=1}^{N} A_i \times sin(2\pi ift + \Phi_i) 
\end{equation}

\noindent where $m$ is the magnitude, $A$ is the amplitude, $f$ 
is the frequency, $t$ is the time of observation, $\Phi_i$ is the phase 
and $i$ runs from 1 to $N$, where $N$ is 9 and 10 for SZ~Tau and RT~Aur, respectively.
The sequence in $N$  was stopped when there were only many low frequency peaks left.   
Tables~\ref{rtfreqs} and \ref{szfreqs} list the result of our Fourier analysis performed 
by using {\tt Period04} (Lenz \& Breger, 2005), i.e. frequencies, amplitudes, and 
phases and their errors for RT~Aur and SZ~Tau. 
Frequencies identified as harmonics of the fundamental oscillation 
are indicated in the first column. Other frequencies found in the 
analysis are listed in the tables (in the order of increasing 
frequency) but are not included in further analysis.  


\begin{table}
\caption{Frequencies, amplitudes and phases of the identified 
   frequency peaks in the frequency spectrum of RT~Aur.}
\label{rtaurfreq}
\begin{tabular}{ccccc}
\hline\hline
ID &   Frequency ($\sigma$) &  Amplitude ($\sigma$)  &  Phase ($\sigma$) & S/N  \\
& d$^{-1}$ & mag & rad & \\
\hline

f$_0$    &     0.26819 (0.00001)  &   0.34690 (0.00009)  &   0.40947 (0.00004)   &   4954  \\
2f$_0$   &    0.53639 (0.00002)  &   0.12895 (0.00009)   &  0.23672 (0.00011)    &  1840 \\
3f$_0$   &    0.80458 (0.00004)  &   0.05576 (0.00009)  &   0.06397 (0.00024)   &    795 \\
4f$_0$   &    1.07277 (0.00009)  &   0.02515 (0.00009)  &   0.92108 (0.00053)   &    357 \\
5f$_0$   &    1.34097 (0.00020)  &   0.01098 (0.00009)  &   0.72003 (0.00123)    &   155 \\
6f$_0$   &    1.60916 (0.00037)  &   0.00582 (0.00009)  &   0.54719 (0.00230)   &     81 \\
7f$_0$   &    1.87736 (0.00075)   &  0.00293 (0.00009)   &  0.37458 (0.00472)   &     40 \\
8f$_0$   &    2.14555 (0.00147)  &   0.00149 (0.00009)   &  0.16564 (0.00917)   &     19 \\
9f$_0$    &   2.41374 (0.00248)  &   0.00078 (0.00009)  &   0.00967 (0.01546)   &      9 \\
10f$_0$   &  2.68194 (0.00344)   &  0.00055 (0.00009)   &  0.81981 (0.02165)    &     6 \\
         &    0.03766 (0.00869)   &  0.00113 (0.00619)  &   0.52539 (0.00839)   &     14 \\
        &     0.08529 (0.00990)   &   0.00103 (0.00505)  &   0.68228 (0.00828)   &     13 \\
        &     0.18232 (0.00210)  &   0.00088 (0.00009)  &   0.63075 (0.01142)    &    11 \\
\hline
\label{rtfreqs}
\end{tabular}
\end{table}


\begin{table}
\caption{Frequencies, amplitudes and phases of the identified 
   frequency peaks in the frequency spectrum of SZ~Tau.}
\label{sztaufreq}
\begin{tabular}{ccccc}
\hline\hline
ID &   Frequency ($\sigma$) &  Amplitude ($\sigma$)  &  Phase ($\sigma$) & S/N \\
& d$^{-1}$ & mag & rad & \\
\hline

f$_0$  &    0.31752 (0.00001) &   0.18800 (0.00001) &   0.44034 (0.00002)  &    6526 \\
2f$_0$ &   0.63307 (0.00010) &   0.00811 (0.00002) &   0.61723 (0.00041)   &     280 \\
3f$_0$ &   0.95463 (0.00023) &   0.00374 (0.00002) &   0.71700 (0.00092)    &    128 \\
4f$_0$ &   1.27259 (0.00057) &   0.00130 (0.00002) &   0.00012 (0.00250)    &    43 \\
5f$_0$  &  1.59586 (0.00077) &   0.00077 (0.00002) &   0.95622 (0.00416)    &      25 \\ 
6f$_0$ &   1.91050 (0.00092) &   0.00062 (0.00002) &   0.39391 (0.00501)    &      20 \\
7f$_0$ &   2.22325 (0.00097) &   0.00059 (0.00002) &   0.50797 (0.00466)   &     19 \\
8f$_0$  &  2.54686 (0.00103) &   0.00054 (0.00002) &   0.94617 (0.00573)    &      17 \\
9f$_0$  &  2.86753 (0.00146) &   0.00038 (0.00002) &   0.15596 (0.00812)    &      11  \\ 
      &    0.04669 (0.00066) &   0.00125 (0.00002) &   0.93945 (0.00309)    &      42 \\
      &    0.18431 (0.00193) &   0.00074 (0.00003) &   0.61063 (0.00441)   &       24 \\
      &    0.24663 (0.00050)  &  0.00275 (0.00003) &   0.53542 (0.00122)   &       94 \\
      &    0.49502 (0.00060) &   0.00168 (0.00002) &   0.92675 (0.00288)    &      57 \\
      &    0.54126 (0.00043) &   0.00243 (0.00002) &   0.74911 (0.00192)    &      83 \\
       &     1.07039 (0.00330) &   0.00113 (0.00014) &   0.44314 (0.00679) &       38 \\
      &    1.11149 (0.00426) &   0.00087 (0.00010) &   0.79006 (0.04108)   &       29 \\
       &   1.14850 (0.00323) &   0.00099 (0.00011) &   0.01984 (0.10169)   &       33 \\ 
      &    1.39226 (0.00074) &   0.00088 (0.00002) &   0.76267 (0.00361)    &      29 \\
\hline
\label{szfreqs}
\end{tabular}
\end{table}


Figures~\ref{rtFourier} and ~\ref{szFourier} show the Fourier spectra of 
RT~Aur and SZ~Tau, respectively. In the left and center panels of Fig.~\ref{rtFourier}, 
the Fourier spectrum of RT~Aur is shown (left) and then again prewhitened by the
10 harmonics in Table~\ref{rtfreqs} (center). A residual orbital signal near
14 c/d is apparent in similar figures which extend out to that frequency because 
the target was interleaved with another
target. The  prewhitened spectrum (Fig.~\ref{rtFourier} center)
shows that the removal of the main frequency and its harmonics reduces
the signal from the star to noise. The right panel of Fig.~\ref{rtFourier} is the window 
function for RT~Aur. Fig.~\ref{szFourier} provides the Fourier spectrum for SZ~Tau.
The prewhitened 
spectrum (Fig.~\ref{szFourier} center)  with the harmonics of the pulsational frequency 
in Table~\ref{rtfreqs} removed, however, leaves a complicated pattern of frequencies. 
This contrasts to the reduction to a noise spectrum in the prewhitened 
RT~Aur spectrum (Fig.~\ref{rtFourier} center).

The noise level and $\sigma$ were estimated from a quiet portion of the 
Fourier spectrum (Figs. .~\ref{rtFourier} and~\ref{szFourier} center). 
  By noise in RT Aur's Fourier spectrum 
we mean the   average of the  
residuals after removing f$_0$ and its harmonics, which we find 
featureless. Sigmas in Tables 1 and 2   are the uncertainty of a given fitted parameter
 determined with the Monte Carlo simulation performed using Period04.
For RT Aur and SZ Tau they are noise: 1.3 $\times$ 10$^{-4}$ and 
4.9 $\times$10$^{-5}$, respectively.
The S/N in Tables 1 and 2 was generated using these values.  



\subsection{Comparison of SZ~Tau and RT~Aur}
 
The RT~Aur data are well represented by Fourier series for a single
frequency and its harmonics and the pattern due to the orbit of
the satellite. The SZ~Tau data, on the other hand, have deviations 
from the Fourier fit.

\subsubsection{Light curve stability}

In  Section 3 we showed that the fundamental pulsator RT~Aur
has the expected behavior for a Cepheid. While there is some scatter
for the moments of maximum brightness, the median brightness
on the ascending branch repeats in a strictly regular manner.
  (Median brightness is the average value of the
adjacent minimum and maximum, which is not identical to 
the mean magnitude calculated for the complete pulsational
cycle).
The other {\it MOST} Cepheid, SZ~Tauri (overtone pulsator) shows
perceptible cycle-to-cycle changes. 
In this section we will consider how this contrasting stability of the
light curves of a fundamental mode and an overtone mode pulsator is
manifested in the Fourier parameters. For this, we fitted a
high-order 
Fourier polynomial at the primary frequency and its harmonics 
using  Equation\ \ref{foupol}   for each cycle. Then, we characterized the 
light-curve shapes with the Fourier parameters (Simon \& Teays 
1982), and         show             results for  
 $R_{21}=A_{2}/A_{1},\ R_{31}=A_{3}/A_{1} and \ R_{51}=A_{5}/A_{1} \ {\rm as \ well \ as}\ \phi_{21}=\phi_{2}- 2\phi_{1}\ {\rm and} \ \phi_{31}=\phi_{3}-3\phi_{1}$.

\begin{figure*}
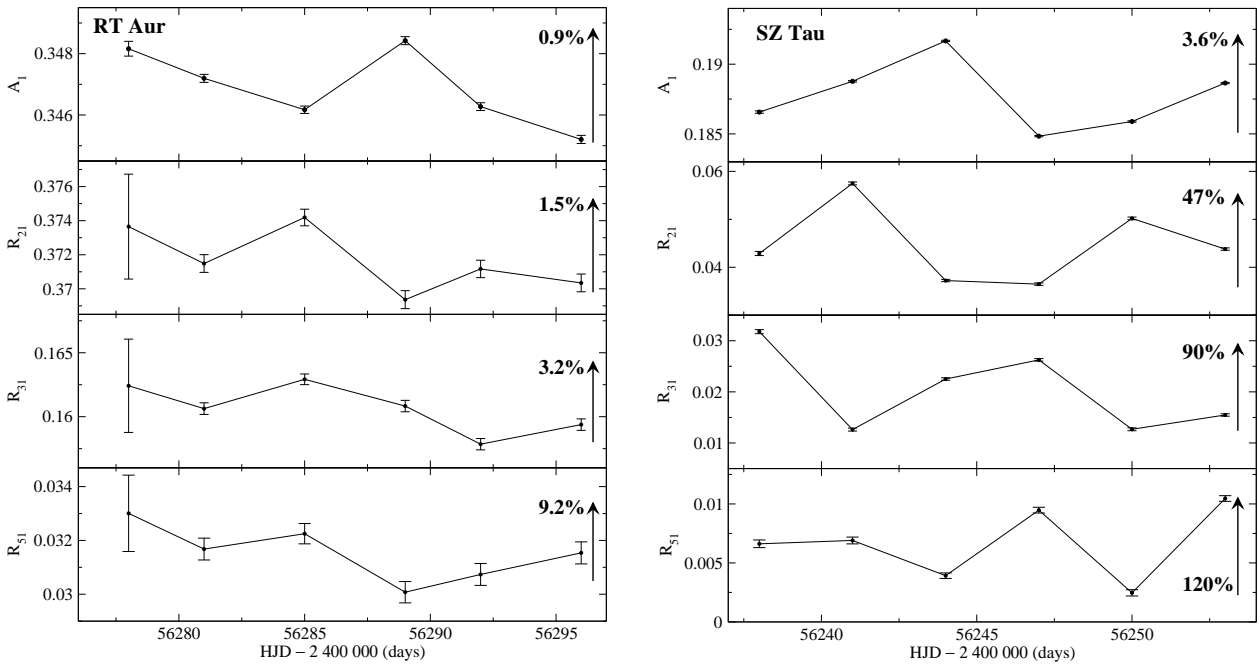

\includegraphics[width=8.0cm]{rtaur_fou_v2.eps}
\hspace{5mm}
\includegraphics[width=8.0cm]{sztau_fou_v2.eps}
\caption{Fourier parameters of RT~Aur (left panels) and SZ~Tau (right panels) from  the {\it MOST} 
observations. The arrows on the right hand side of each panel show the approximate 
variation of the parameters during the cycles observed; the percentage of variation 
is given by each arrow. 
 Top panels: variation of the $A_1$ amplitude between cycles. The panels below show
the  variation of the amplitude ratios    $R_{21}$,  $R_{31}$, and $R_{51}$.   The errors in 
the first cycle of RT Aur are unusually large because the cycle was incompletely covered.}
\label{fouamp}
\end{figure*}

\hfill
\vfill
\eject

\begin{figure*}
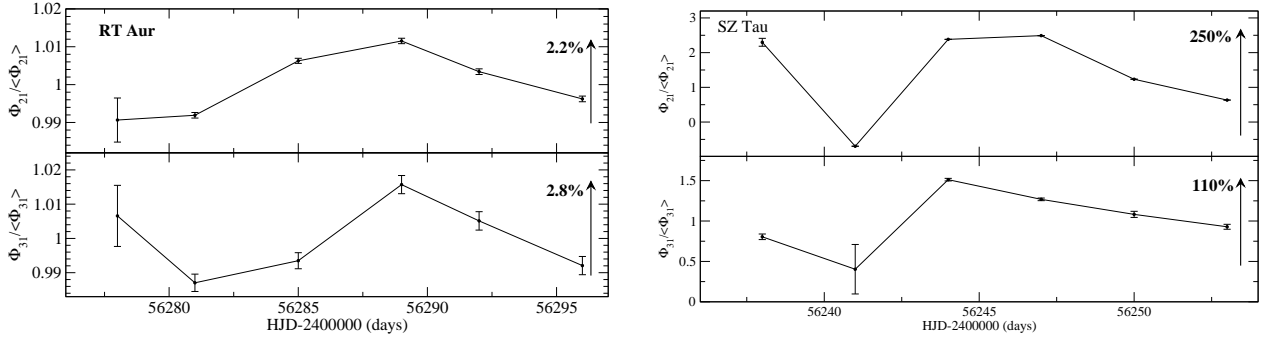

\includegraphics[width=8.0cm]{rtaur_fi.eps}
\hspace{5mm}
\includegraphics[width=8.0cm]{sztau_fi.eps}
\caption{Fourier parameters of RT~Aur (left panels) and SZ~Tau (right panels) from  the {\it MOST} 
observations. Again, the arrows indicate the approximate variation between cycles.  
Top panels: the variation of ${\phi}_{21}/{\langle}{\phi}_{21}{\rangle}$. Bottom panels: variation of    ${\phi}_{31}/{\langle}{\phi}_{31}{\rangle}$.   The errors in 
the first cycle of RT Aur are unusually large because the cycle was incompletely covered.  }
\label{fouphi}
\end{figure*}

\begin{figure*}
\includegraphics[angle=270,width=7.5cm]{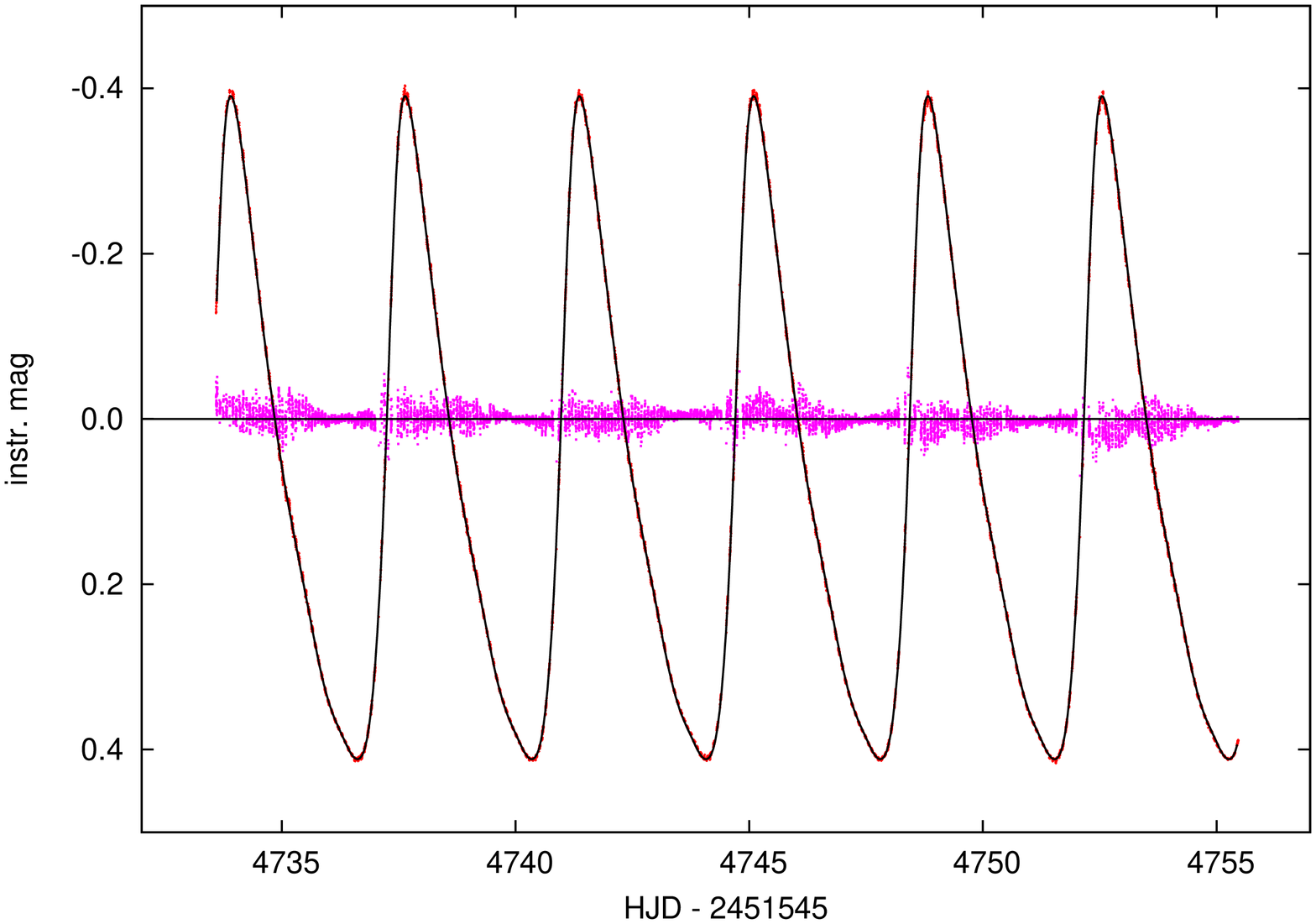}
\includegraphics[angle=270,width=7.5cm]{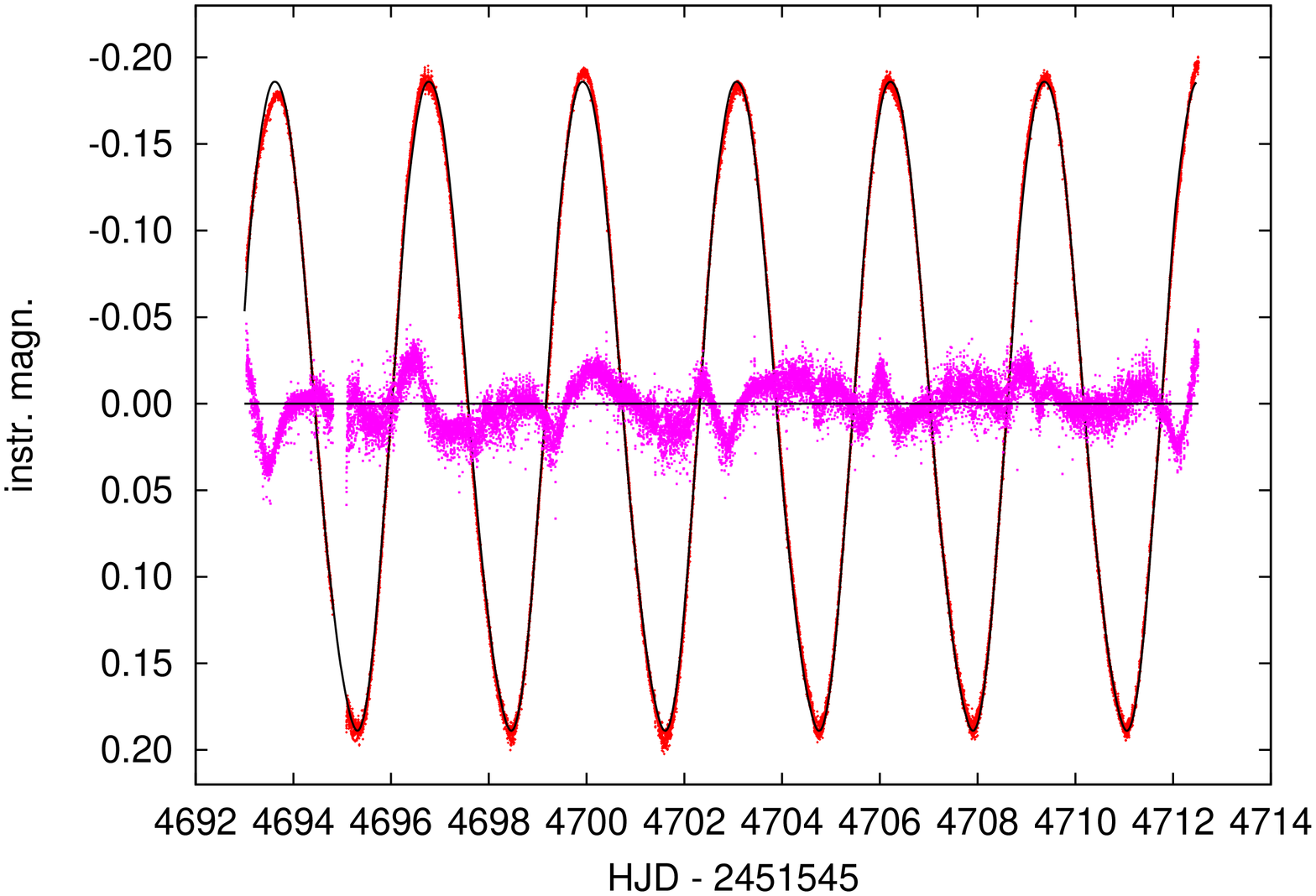}\\
\includegraphics[angle=270,width=7.5cm]{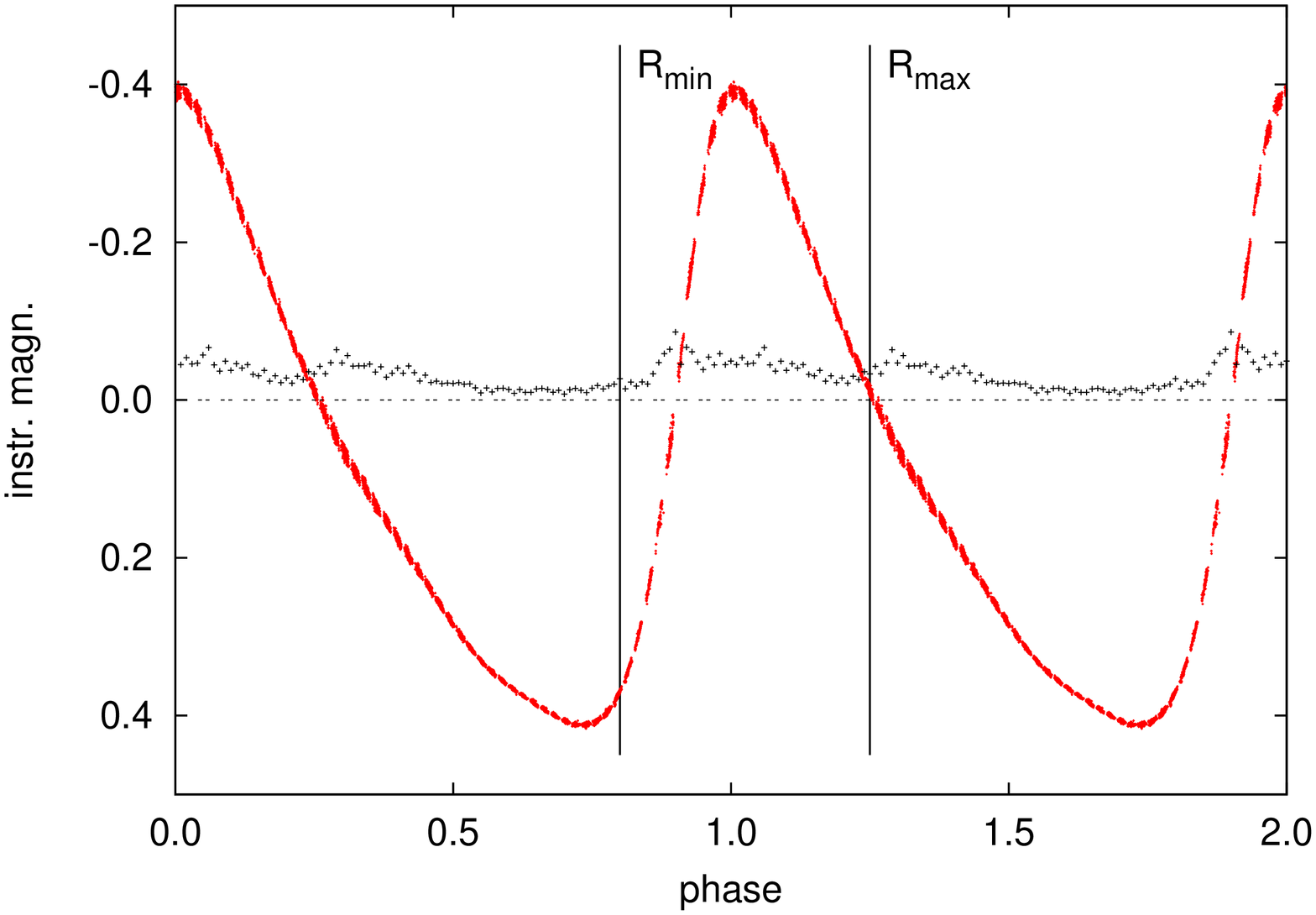}
\includegraphics[angle=270,width=7.5cm]{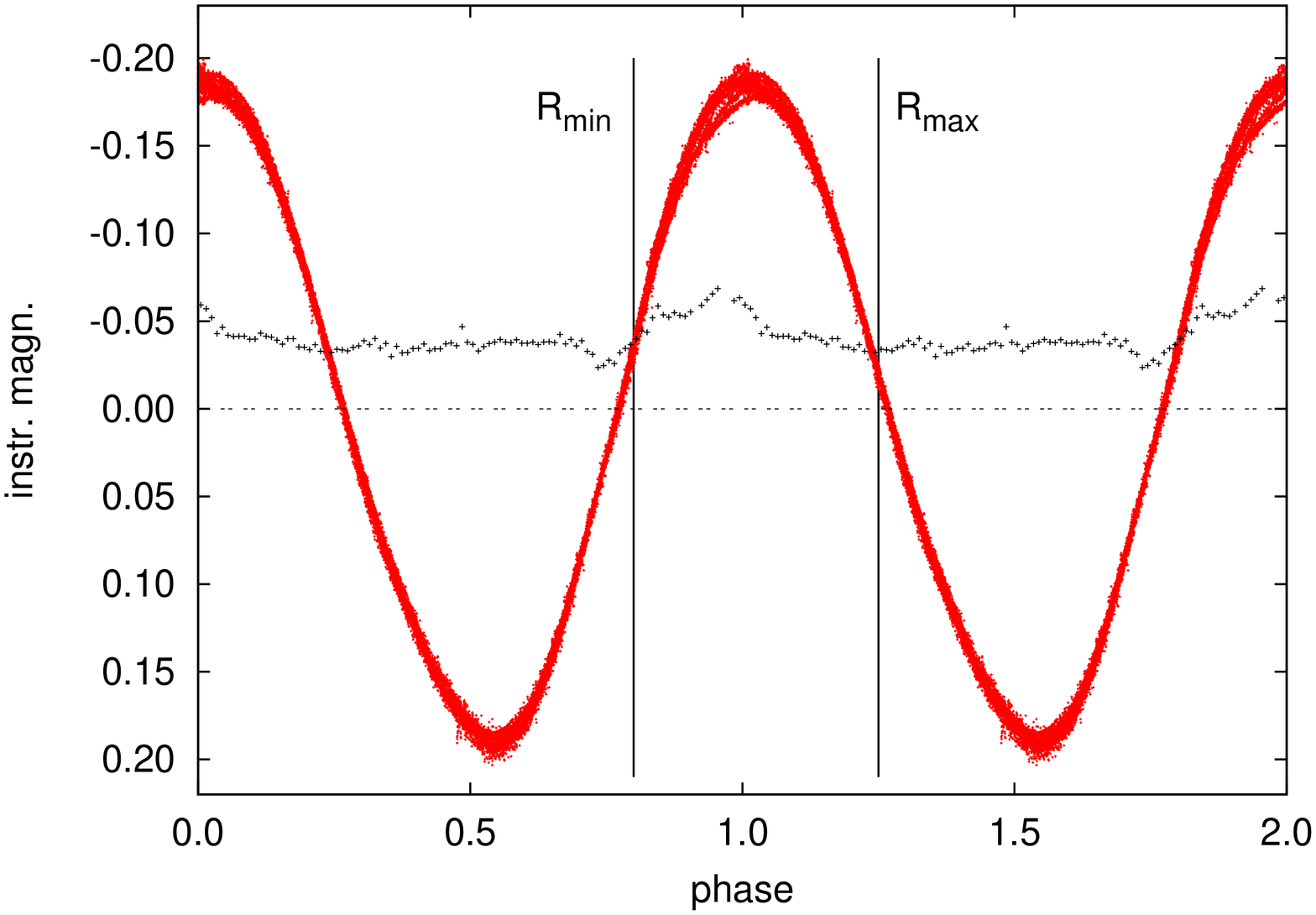}
\caption{The fit of the light curve of RT~Aur (top left panel) and SZ~Tau
  (top right panel). 
Red points: observations, black curve: Fourier fit, magenta
points: residuals magnified 3 times to enhance 
visibility. Residuals were created from  a 10-term Fourier fit of
RT~Aur from
Table~\ref{rtfreqs} (left panel) and 
 for SZ Tau  from  a 9-term Fourier fit taken from
Table~\ref{szfreqs} (right panel). Bottom: Standard deviations between
cycles as a
function of phase binned into 100 phase bins.  Left: RT Aur: The phased 
light curve (red) is
compared with the standard deviations between cycles, magnified by 3
 times for plotting (black dots). Maximum and minimum radius are also shown.
Right: The same for SZ Tau.   }
\label{sztau_fit}
\end{figure*}

The top left panel of Fig.~\ref{fouamp} shows the amplitude $A_1$ of RT~Aur for the 6 cycles covered. The arrows in each panel show the approximate range of variation between 
cycles, together with the percentage of variation.  
Note that the variation in this amplitude is only about  0.9\% between the cycles.  
In SZ~Tau (top right panel in Fig.~\ref{fouamp}) the variation in the
Fourier amplitude $A_1$ is  3.6\%. Similarly, the variation of the  amplitudes 
    $R_{21}$,  $R_{31}$, and  $R_{51}$
is shown in Fig.~\ref{fouamp} for RT~Aur (lower left panels) and for SZ~Tau (lower  right panels). For RT~Aur    $R_{21}$, $R_{31}$, and  $R_{51}$
vary by approximately 1.5\%, 3.2\%, and 9.2\% respectively. For SZ~Tau, the respective
variations are closer to 50\%,  100\% and 120\%. Similarly, the variations
in phase parameters
 $\phi_{21}$ and $\phi_{31}$ are modest in RT~Aur (left panels in Fig.~\ref{fouphi}) but
much larger in SZ~Tau (right panels in Fig.~\ref{fouphi}).  

In summary, all the light curve quantities are much more variable in 
SZ~Tau than in RT~Aur. We can also compare the variations in RT~Aur with 
those of the fundamental mode Cepheid in the Kepler field (V1154~Cyg), 
which has a much longer series of observations (Derekas et~al. 2012).
The $A_1$ amplitude for V1154~Cyg has comparably small variations over 
the intervals that correspond to the length of the {\it MOST} observations. 
However, on time scales as long as a year it has larger amplitude variations. 
Thus there may be larger variations than we have seen in the {\it MOST} RT~Aur data, but
it is on a much longer time scale. Other Fourier parameters in V1154~Cyg 
have variations comparable to those of RT~Aur.

\subsubsection{Phase dependence}

 To further study the deviations from the Fourier 
representation of SZ~Tau, 
Fig.~\ref{sztau_fit} (right) shows the deviations from the 9-term representation.  
  The high quality of the data allows us to inspect individual cycles of the two stars.  
As a comparison, 
the left panel of Fig.~\ref{sztau_fit} features the much smaller residuals of RT~Aur. These small
residuals show  clear pattern however. They almost vanish around
minimum light, and are larger in other pulsational phases.  For SZ Tau, there
are some differences in the residuals from cycle to cycle (Fig.~\ref{sztau_fit} top
right).  However, 
the deviations have a reasonably consistent pattern,
including sharp inflections following minimum radius 
(Fig.~\ref{sztau_fit}, top right).
 When the standard deviations between cycles are created as a
function of phase (Fig.~\ref{sztau_fit} bottom), the phase dependence is 
surprisingly similar for RT
Aur (left) and SZ Tau (right).  That is both the fundamental and
overtone pulsators show an increased standard deviation around minimum
radius when the next pulsation cycle is initiated by a ``shove'' from
the pulsation piston.  After maximum radius, the ``coasting phase''
the deviations between cycles decrease, and remain small until the
``shove'' from the next cycle.

\section{{\boldmath $O-C$} diagrams}

\begin{figure*}
\includegraphics[width=8cm]{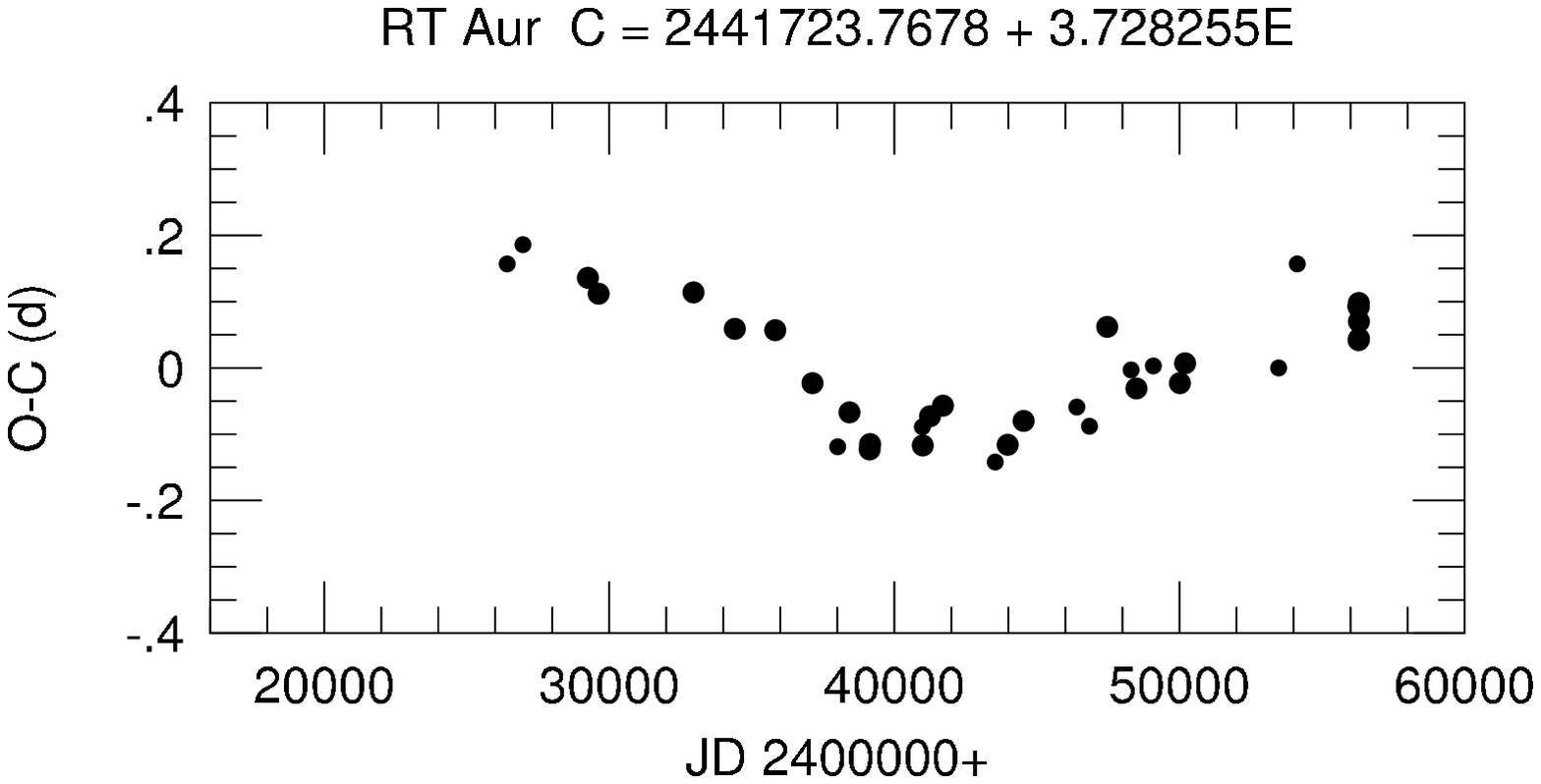}\includegraphics[width=8.2cm]{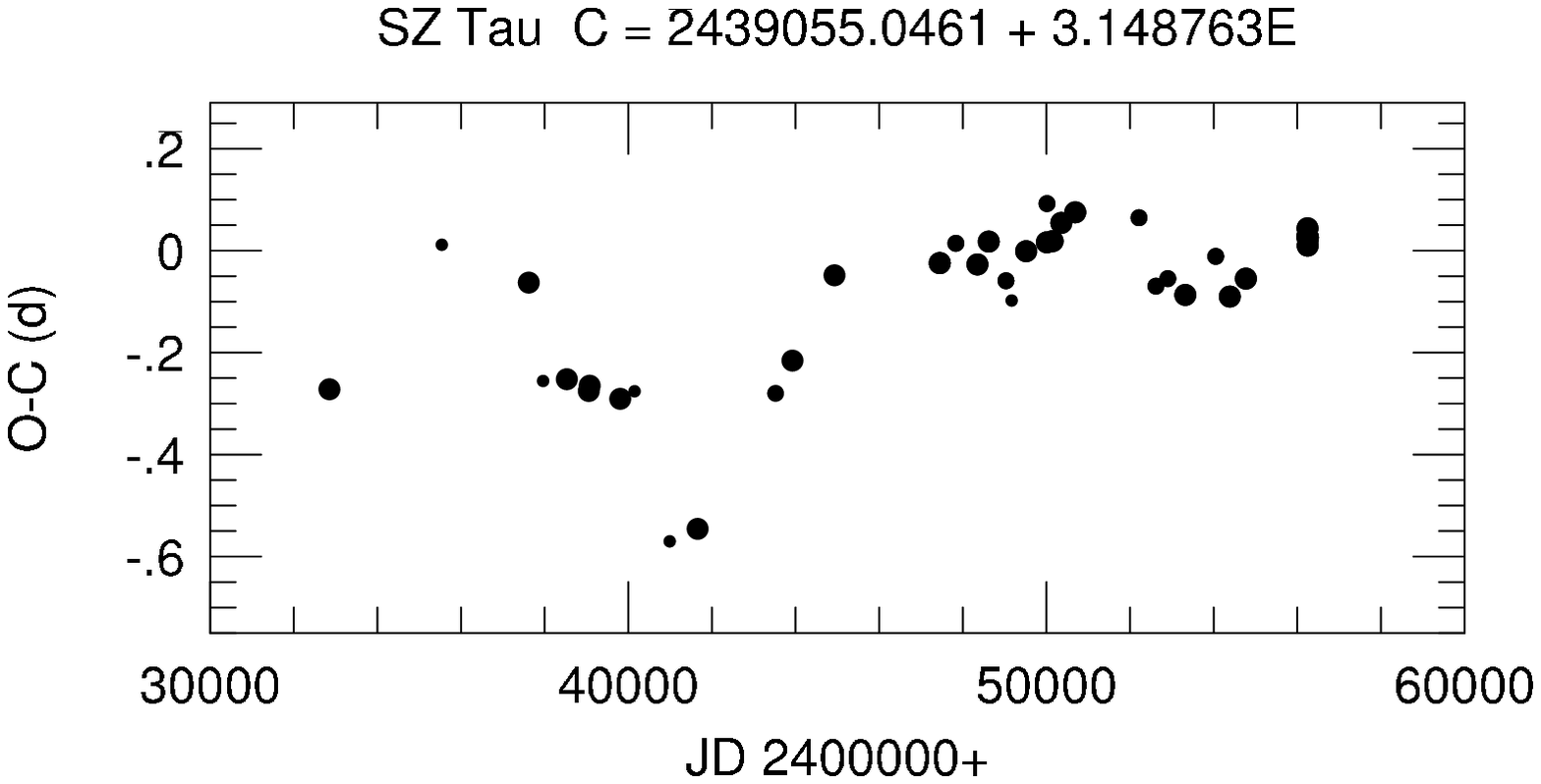}
\caption{Long-term $O-C$ variations of RT~Aur (left panel) and SZ~Tau 
(right panel) with the epoch and period used. While both stars show 
significant variations, RT~Aur seems to change its pulsational period 
more smoothly than SZ~Tau. 
  Symbol sizes correspond to the weight assigned to the 
$O-C$ value listed in Tables~7 and 8.}
\label{rt-omc}
\end{figure*}

\begin{figure*}
\includegraphics[width=8.2cm]{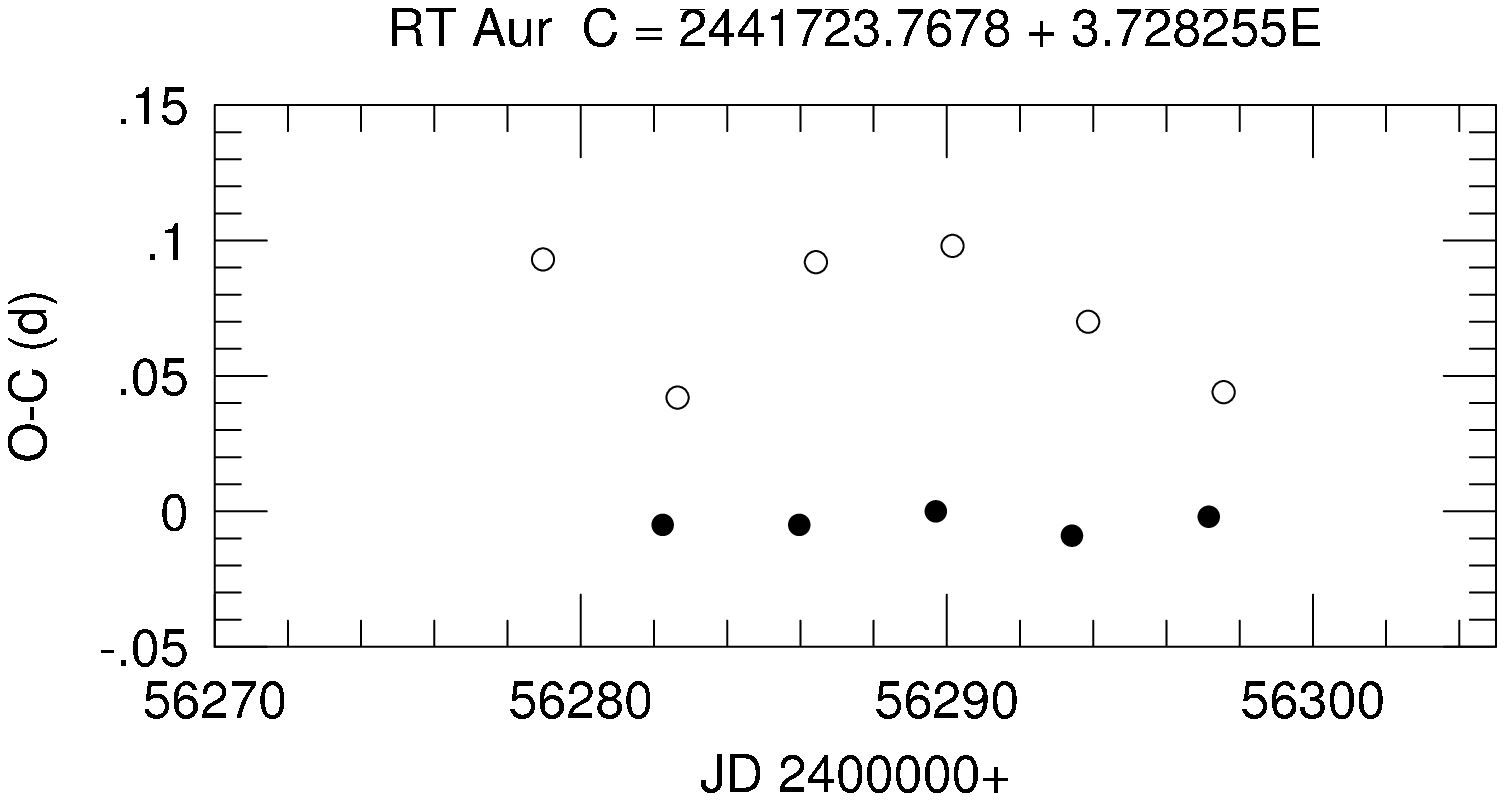}\includegraphics[width=8.2cm]{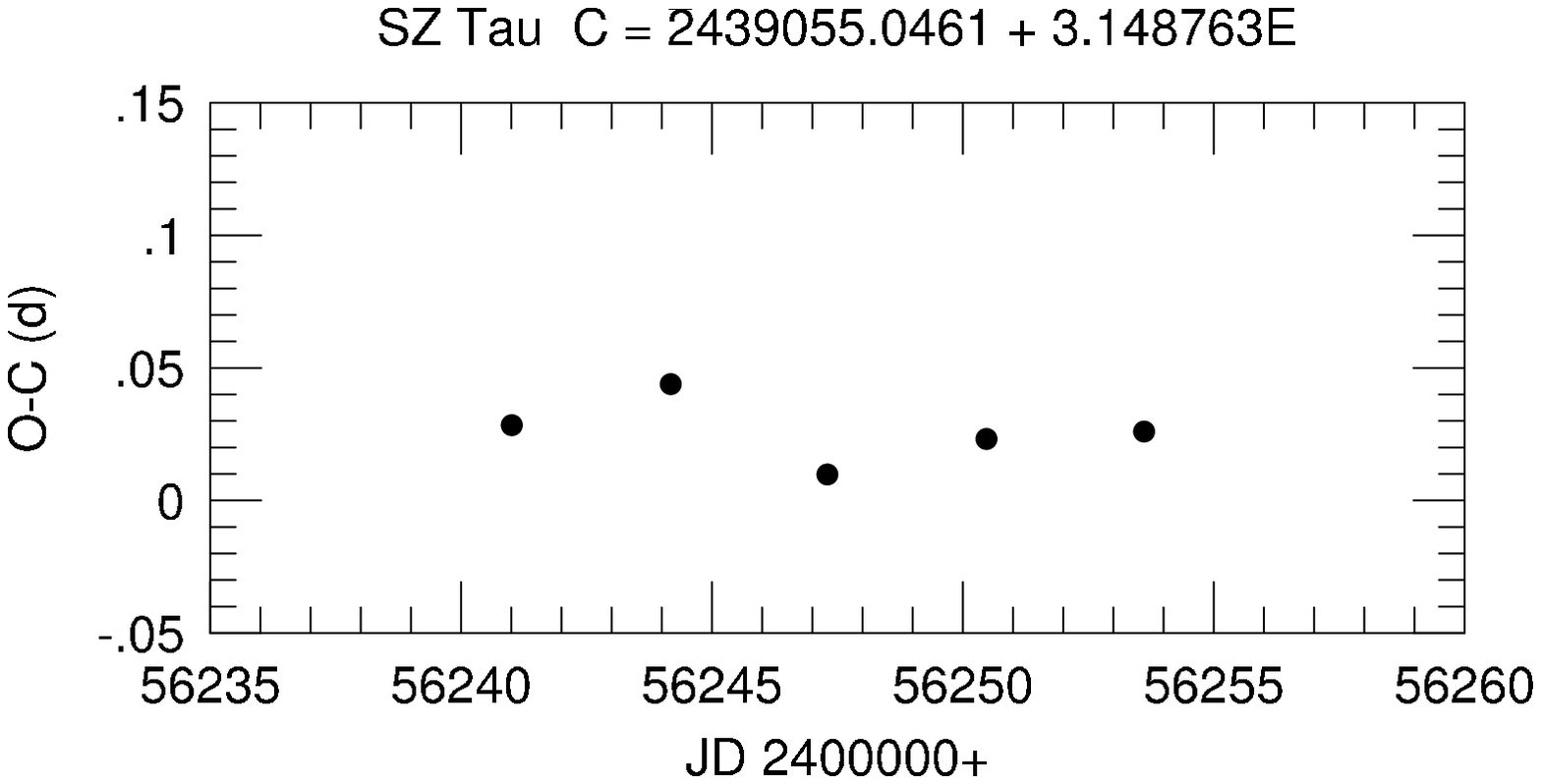}
\caption{Left: $O-C$ variations of RT~Aur based on the {\it MOST}
 observations. Open circles refer to moments of brightness maxima, filled circles
 represent $O-C$ values for moments of the median brightness on the
 ascending branch (offset by 0.360 day to facilitate a direct comparison 
 with those for light maxima) Right: $O-C$ variations of SZ~Tau based on the {\it MOST} 
 observations.}
\label{rt-mostoc}
\end{figure*}



To put the {\it MOST} observations into the long-term context, Fig.~\ref{rt-omc}
shows the $O-C$ diagrams of both stars and Fig.~\ref{rt-mostoc} shows the $O-C$ diagrams 
from the {\it MOST} observations.

\subsection{RT~Aur}

For RT~Aur, several studies of the variations in the pulsation period 
are available in the literature, of which the latest are: 
Szabados (1991), Berdnikov et~al. (2003), Meyer (2004), and Turner et~al. 
(2007). The $O-C$ diagram based on the normal brightness maxima published 
by Turner et~al. (2007) is shown in the left panel of Fig.~\ref{rt-omc}. 
Their Tables 2 and~3 are used, 
but data with an assigned weight less than 1 have been omitted  and 
several values based on photometric data which they omitted are included, 
as well as the new {\it MOST} data. The new set of normal maxima together 
with references is listed in Table~\ref{tab-rtaur-oc}, which is available 
in the electronic version. In that table the $O-C$ residuals correspond to 
the updated ephemeris:
\begin{equation}
C_{\rm max} = 2\,441\,723.7678 + 3.728\,255{\times}E 
\label{rtaur-ephemeris}
\end{equation}
\vspace{-3mm}
$\phantom{mmmmmmml}\pm0.0077\phantom{}\pm 0.000\,003$

This ephemeris was obtained by a weighted linear least
squares method based on the tabulated normal maxima.
A weight of 1, 2 , or 3 has been assigned to the individual data 
series depending on the number and quality of the data points. 
For a better visualization, we only
plotted the most reliable $O-C$ residuals (those with
weights 2 or 3) in the left panel of Fig.~\ref{rt-omc}. Part
of the scatter may be due to the 
uncertainty in the determination of the moments of 
individual normal maxima. However, another cause is
also present. The 0.05 day difference within 3 weeks
between the $O-C$ values for the brightness maxima 
(denoted as open circles in the left panel of Fig.~\ref{rt-mostoc})
from the accurate and uninterrupted {\it MOST} data  
is intrinsic to stellar pulsation. It is also seen, however, 
that the pulsation period
of RT~Aur is very stable for the moments of the median
brightness on the ascending branch of the light curve (filled
circles in the left panel of Fig.~\ref{rt-mostoc}).
  This means that the fluctuation in the $O-C$ values is caused by the
slight variability in the shape of the light curve. In fact, the 0.05
day fluctuation in the {\it MOST} $O-C$ values corresponds to 1.3\%
of the pulsation period which is quite compatible with the fluctuations
of the Fourier amplitude and phase parameters visualized in the left
panel of Fig.~6.



\begin{table}
\caption{$O-C$ values of RT~Aurigae. $E$ is the epoch of the
  determination in cycles; $W$ is the weight. This is only a portion of the 
full version available online only.}
\begin{tabular}{l@{\hskip2mm}r@{\hskip2mm}r@{\hskip2mm}c@{\hskip2mm}l}
\hline
\noalign{\vskip 0.2mm}
JD$_{\odot}$ & $E\ $ & $O-C$ & $W$ & Data source \\
2\,400\,000 &  & (d)\,\,\,\, && \\
\noalign{\vskip 0.2mm}
\hline
\noalign{\vskip 0.2mm}
 20957.478 & $-$5570 & 0.091 & 1 & Turner et~al. (2007) \\
 22784.241 & $-$5080 & 0.009 & 1 & Turner et~al. (2007) \\
 26419.438 & $-$4105 & 0.157 & 2 & Turner et~al. (2007) \\
 26971.249 & $-$3957 & 0.186 & 2 & Turner et~al. (2007) \\
 27504.400 & $-$3814 & 0.197 & 1 & Turner et~al. (2007) \\
\noalign{\vskip 0.2mm}
\hline
\end{tabular}
\label{tab-rtaur-oc}
\end{table}

\begin{figure}
\includegraphics[width=8cm]{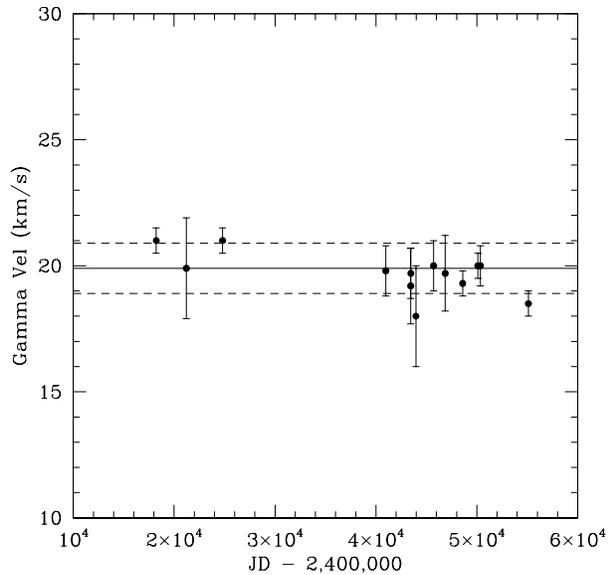}
\caption{Variations 
of the $\gamma$ (systemic) velocity of RT~Aur. The solid line is the
mean; dashed lines indicate $\pm$ 1 km\,s$^{-1}$ from the mean. None
of the velocities deviate from the $\pm$ 1 km\,s$^{-1}$ band by more
than 1 $\sigma$. }
\label{gamma}
\end{figure}

In RT~Aur the $O-C$ diagram (period change; left panel of
Fig.~\ref{rt-omc}) has relatively 
smooth variation.   If the pattern of the $O-C$ residuals is approximated 
by a positive parabola, then the  
 rate of period increase is 0.000986~d/century.
The pattern of the $O-C$ residuals of RT Aur  has been interpreted in
different ways, from 
a simple parabola (Fernie 1993) to a more complicated form (Turner et~al. 
2007). The amplitude from the $O-C$ variations, however, is commensurate with 
an evolutionary time scale, and as noted by Fernie, is much smaller than 
that of the first overtone pulsator Polaris. Turner et~al. interpret a 
wavelike structure to the $O-C$ residuals as light-time effect in a very 
long period orbit (26429 days = 72 years); however, they find this 
difficult to reconcile with the constraints on the companion.  
We present a summary of the systemic ($\gamma$) velocity data in Fig.~\ref{gamma} 
and Table~\ref{tab-rtaur-vgamma} including the most recent data from 
Takeda et~al. (2013).   
None of the systemic velocities differs from a band of 
$\pm 1$ km\,s$^{-1}$ around the mean value by more than 1~$\sigma$, 
making it unlikely that orbital motion has been detected, although
a tendency of a slight decrease in the $\gamma$-velocity values is
noticeable. Further observations are necessary to clarify the question
of binarity of RT~Aur.



\begin{table}
\caption{$\gamma$-velocities of RT~Aurigae.}
\begin{tabular}{l@{\hskip2mm}c@{\hskip2mm}c@{\hskip2mm}l}
\hline
\noalign{\vskip 0.2mm}
JD & $v_\gamma$ & $\sigma$ & Data source\\
2\,400\,000 & (km\,s$^{-1}$) & (km\,s$^{-1}$) &\\
\noalign{\vskip 0.2mm}
\hline
\noalign{\vskip 0.2mm}
18230  & 21.0 &  0.5 &  Petrie (1932) (obs.: Duncan)\\
21210  & 19.9 &  2.0 &  Kiess (1917)\\
24800  & 21.0 &  0.5 &  Petrie (1932)\\
40979  & 19.8 &  1.0 &  Evans (1976)\\
43449  & 19.2 &  1.5 &  Wilson et~al. (1989)\\
43457  & 19.7 &  1.0 &  Beavers \& Eitter (1986)\\
43962  & 18.0 &  2.0 &  Barnes et~al. (1987)\\
45717  & 20.0 &  1.0 &  Gieren (1985)\\
46866  & 19.7 &  1.5 &  Gorynya et~al. (1996)\\
48600  & 19.3 &  0.5 &  Gorynya et~al. (1996)\\
50100  & 20.0 &  0.5 &  Gorynya et~al. (1996)\\
50350  & 20.0 &  0.8 &  Kiss (1998)\\
55110  & 18.5 &  0.5 &  Takeda et~al. (2013)\\
\noalign{\vskip 0.2mm}
\hline
\end{tabular}
\label{tab-rtaur-vgamma}
\end{table}

\subsection{SZ~Tau}

SZ~Tau, on the other hand, has more erratic variations in its $O-C$
diagram,    and in 
particular, times when the $O-C$ is increasing alternating with 
periods when it is decreasing. That is, periods are neither monotonically 
increasing nor decreasing as expected for evolution through the 
instability strip. $O-C$ diagrams showing the variability of the 
pulsation period for many stars are available in Szabados (1977, 1991) and Berdnikov 
\& Pastukhova (1995). These latter authors approximated the $O-C$ 
graph of SZ Tau with a parabola implying a continuous period decrease with 
erratic changes superimposed. According to Szabados (1977, 1991), 
however, the linear sections in the $O-C$ graph imply that there are 
preferred values of the period to which the Cepheid returns during 
its pulsation on a time scale of several years-decades. The $O-C$ 
residuals based on previous photoelectric and CCD observations are plotted 
in the right panel of Fig.~\ref{rt-omc}, extending the time base by two decades
as compared with the latest one by Berdnikov \& Pastukhova (1995).

The $O-C$ residuals based on all available accurate (photoelectric
or CCD) measurements are listed in Table~\ref{tab-sztau-oc} together
with the references in the electronic version.
  For low-amplitude Cepheids, such as SZ~Tau, the preferred light-curve
feature  for $O-C$ studies is not the moment of the
light maximum because of the large uncertainty in determining the
phase of the brightness extremum during the shallow variation.
Instead, the behaviour of the pulsation period can be followed by
studying the moment of the median brightness on the ascending branch
of the light curve where the variations in the brightness are the
steepest during the pulsational cycle (see Sect.~2.2 in Derekas et~al.
2012).
The residuals are determined from the $V$ band light curves 
(or the nearest band to it), and refer to the moments of median 
brightness on the ascending branch corresponding to the ephemeris:
\begin{equation}
C_{\rm med} = 2\,439\,055.0461 + 3.148\,763{\times}E 
\label{sztau-ephemeris}
\end{equation}
\vspace{-3mm}
$\phantom{mmmmmmml}\pm 0.0283 \phantom{}\pm 0.000\,006$

A least squares linear fit has been applied to the $O-C$ residuals 
after $E=2666$ = HJD 2,447,450 that resulted in Eq.~\ref{sztau-ephemeris}. The 
resulting $O-C$ diagram is plotted in the right panel of Fig.~\ref{rt-omc}.
Table~\ref{tab-sztau-oc} contains some $O-C$ residuals based on 
photoelectric $UBV$ observations obtained with the photometer attached 
to the 50~cm Cassegrain telescope at the Piszk\'estet\H{o} Mountain 
Station of the Konkoly Observatory between 1991 and 2003. The individual 
photometric data are listed in Table~\ref{tab-sztau-newdata} 
(observer: L.~Szabados).  The full table is provided in the electronic
version. 

\begin{table}
\caption{$O-C$ values of SZ~Tauri. This is only a portion of the 
full version available online only.}
\begin{tabular}{l@{\hskip2mm}r@{\hskip2mm}r@{\hskip2mm}c@{\hskip2mm}l}
\hline
\noalign{\vskip 0.2mm}
JD$_{\odot}$ & $E\ $ & $O-C$ & $W$ & Data source\\
2\,400\,000 + && (d)\,\, &&\\
\noalign{\vskip 0.2mm}
\hline
\noalign{\vskip 0.2mm}
32851.7111 & $-$1970 & $-$0.2719 & 3 & Eggen (1951)\\
35541.0382 & $-$1116 &   0.0116 & 1 & Walraven et~al. (1958)\\
37619.1476 & $-$456 & $-$0.0626 & 3 & Mitchell et~al. (1964)\\
37962.1697 & $-$347 & $-$0.2556 & 1 & Williams (1966)\\
38528.9506 & $-$167 & $-$0.2521 & 3 & Wisniewski \& Johnson (1968)\\
\noalign{\vskip 0.2mm}
\hline
\end{tabular}
\label{tab-sztau-oc}
\end{table}


\begin{table}
\caption{New $UBV$ photometric data of SZ~Tauri. This is only a 
portion of the full version available online only.}
\begin{tabular}{l@{\hskip2mm}c@{\hskip2mm}c@{\hskip2mm}c}
\hline
\noalign{\vskip 0.2mm}
JD$_{\odot}$ & $V$ & $B-V$ & $U-B$\\
2\,400\,000 + & (mag) & (mag) & (mag)\\
\noalign{\vskip 0.2mm}
\hline
\noalign{\vskip 0.2mm}
48593.5264 & 6.385 & 0.739 & 0.526\\
48594.5173 & 6.616 & 0.874 & 0.612\\
48646.3869 & 6.488 & 0.779 & 0.528\\
48647.3421 & 6.431 & 0.758 & 0.562\\
49032.3006 & 6.670 & 0.875 & 0.615\\
\noalign{\vskip 0.2mm}
\hline
\end{tabular}
\label{tab-sztau-newdata}
\end{table}

The scatter of the plot for the last decades where the pulsation 
period was approximated as a constant value of 3.148763~d exceeds 
the observational uncertainty, and this is the case for the $O-C$ 
residuals derived from the {\it MOST} data, as well. Cycle-to-cycle 
period change is present in the right panel of Fig.~\ref{rt-mostoc}, 
similar to V1154~Cyg. 
The period in individual cycles varies within 0.7\%  of the 
average value of the pulsation period. On a longer time scale, there 
are erratic cycle-to-cycle variations (right panel of Fig.~\ref{rt-omc}).


\subsection{Pulsation mode}

  The erratic cycle-to-cycle variation of SZ~Tau (as compared with RT~Aur) 
is characteristic of overtone pulsators (e.g. Fig.~2 in Berdnikov 
et~al. 1997). In the past shorter data strings have been fitted to 
parabolas implying to rapid period changes (see Szabados 1983) 
but more extensive data appear to suggest that for many overtone 
pulsators the period variations are more complicated than simple 
monotonic changes.

  The rigorous interpretation of $O-C$ diagrams has been discussed extensively by 
Koen (2006), based on the combination of measurement error, a long term change in period, and
random changes in the period.  Specifically, he demonstrates that it is possible for 
random changes in the period to mimic long term changes in the period.  Full analysis of 
the $O-C$ period changes in Figure~\ref{rt-omc} is beyond the scope of this paper, but we will
put the $O-C$ charactistics of the overtone (SZ Tau) in context, and discuss instances where
the $O-C$ diagram is consistent with a Koen dominant long term variation, and where 
it is consistent with a Koen dominant random period jitter.

In this section, we develop a    qualitative   summary  of period
variation  in overtone Cepheids based on the $O-C$ curves of Berdnikov
et~al. (1997) of low amplitude Cepheids.  
The first step is to confirm the pulsation mode.  For this
we have used  primarily the classifications 
of Groenewegen \& Oudmaijer (2000), Kienzle et~al. 
(1999) and Sachkov (1997). The pulsation mode of V1334~Cyg was discussed 
by Evans (2000) and the pulsation mode of V473~Lyr by Burki et~al. 
(1986). FF~Aql was found not to be an overtone pulsator by Benedict et~al. 
(2007). In 3 cases, (GI~Car, V532~Cyg, and VZ~CMa) the star was classified
as an overtone pulsator by Kienzle et~al. but not Groenewegen \& Oudmaijer. 
We have retained the overtone designation since the sensitivity to
pulsation mode varies  depending on both  the  period ranges 
and the diagnostic itself. 

We then have used the  $O-C$ diagrams of 
 Berdnikov et~al., and identified three categories of period change.  
First, 7 stars (EU~Tau, $\alpha$~UMi, 
SU~Cas, GI~Car, V1726~Cyg, V473~Lyr [second overtone pulsator],
and UY~Mon) were found to have $O-C$ diagrams consistent with a parabolic
fit (Figs. 1 and 2 in Berdnikov et~al.) all indicating increasing
period (discussed below). Only EU~Tau and $\alpha$~UMi, however, have $O-C$
diagrams which were unquestionably parabolic fits, showing period variation 
which is particularly rapid.   These are consistent with the Koen class of long 
term period variation.   

In the second
group, 7  other stars 
(BY~Cas, V379~Cas, DT~Cyg, V532~Cyg, DX~Gem, EV~Sct, and SZ~Tau; Figs.~3
and 4 in Berdnikov et~al.) clearly have variations in their $O-C$ diagrams, 
but the variations switched from between positive and negative and back 
in an apparently cyclic way. SZ~Tau (right panel of Fig.~\ref{rt-omc}) exhibits 
this behavior. This ``activity'' in the period is clearly not caused
by monotonic evolution 
through the instability strip,   and suggest a that Koen period jitter is 
dominant.   The excursions around a mean period suggest a 
pulsation related cause rather than a secular change due to evolution. 

There is a third group of 15 stars (VZ~CMa, 
GH~Car, V419~Cen, BG~Cru, V1334~Cyg, V526~Mon, QZ~Nor, V440~Per, EK~Pup, 
MY~Pup, V335~Pup, V950~Sco, AH~Vel, FZ~Car, and AZ~Cen; Fig.~5 in Berdnikov 
et~al.) which have no pattern in the photometric $O-C$ residuals. 
This may be because the data series are not long enough, or the 
values are not accurate enough or in many cases the gaps between the times 
of observation obscure any correlation. It is likely that at least some 
of these have period   jitter, possibly at a lower level and over a 
longer time scale than the more prominent fluctuations in the first two 
groups. 

In summary, the photometric monitoring of these 29 overtone 
pulsators shows that nearly half of them  have period 
variation. This includes the second group showing 
substantial  period   jitter.  
Thus on the scale of decades there is evidence of considerable period variation
in overtone pulsators,   including both the Koen categories of long term change
and random period jitter.  


\section{Discussion}

 The {\it MOST} observations display two
characteristics of pulsation not seen before in less plentiful and
less accurate data. First, the light curves as exhibited by the 
Fourier parameters are more variable in the overtone pulsator SZ Tau
than in the fundamental mode pulsator RT Aur. On the other hand, the
differences between cycles display a similar pattern as a function of
phase (Fig. 6). We have made preliminary explorations for explanations
as discussed in the sections below.

\subsection{Mode dependence}

The intensive observations of RT~Aur and SZ~Tau with the {\it MOST}
satellite have demonstrated that the overtone pulsator (SZ~Tau) has more
instability in its pulsation cycle than the fundamental mode Cepheid 
(RT~Aur). This is apparent in simple repetition of the light curve 
from cycle to cycle (Figs. 1 and 2), the Fourier spectrum of the
observations (Figs. 3 and 4), the 
Fourier parameters cycle by cycle (Figs. 5 and 6) and the $O-C$
diagrams (Figs. 8 and 9). As a preliminary consideration, we note that
the node  of a first overtone pulsator occurs higher in the envelope
than that of a fundamental mode pulsator.  This may create differences
in the pulsation even for stars of reasonably similar period, mass,
and temperature.

\subsection{Effect of turbulent convection} 

The quality and quantity of
{\it MOST} and {\it Kepler} satellite observations has revealed
changes in the periods and light curves of SZ~Tau and
V1154~Cyg. Explanations due to evolution, mass loss, and binary
light-time effect are not adequate for the nonmonotonic variations as
discussed in Section 5.3.
It is suggested that there is
an instability in the pulsation itself which is responsible.  

In order to check whether turbulent convection in the partial
ionization  zones can cause cycle-to-cycle light curve variations  
we computed models with 
the Florida-Budapest code (Koll\'ath et al. 2002). We used similar 
parameters and settings as Szab\'o, Buchler \& Bartee (2007). To  
study SZ Tau we computed an overtone  pulsator model, with a mass 
of 5.25 M$_{\odot}$, 
$L= {\rm 1979~L_{\odot}}$ and $T_{\rm eff}$ = 6075 K and solar 
metallicity. The model has a linear overtone
period of 3.106 days. 

\begin{figure}
\includegraphics[width=15cm,bb=-756 50 554 770, clip=false, angle =270]{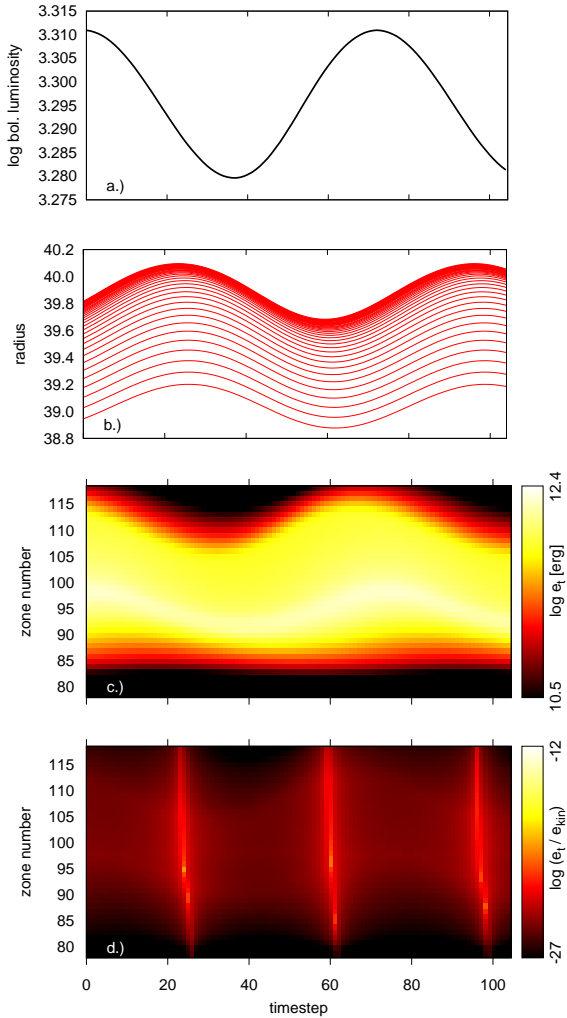}
\caption{a) Bolometric luminosity variation of our model overtone Cepheid;  
b) Radius variation of upper zones in the model; 
c) Turbulent energy as a function of time and zone number; 
d) Logarithm of the ratio of the turbulent and kinetic energy for each zone as a function of time.
Maximum values are reached when local  velocity approaches its minima. }
\label{model}
\end{figure}

Figures \ref{model}a and \ref{model}b show the variation of the bolometric luminosity and the radii 
of the topmost layers of the model after reaching its limit cycle corresponding to the 
radial first overtone mode, respectively. In panel c we plot the turbulent energy as a function of 
time and zone number (from 78 to 121). The magnitude of the turbulent energy is color coded.
This range of zones captures the hydrogen ionization zone featuring the strongest turbulent 
energy throughout the pulsational cycle, on the order of $10^{10} - 10^{12}$ erg per zone. The 
energy is highest where the {\it MOST} light curve of SZ Tau shows the largest scatter. 

However, the magnitude of the turbulent energy content is negligible compared to the kinetic 
energy due to the pulsation as is shown in Fig.~\ref{model}d. Here we show the logarithm of
the ratio of the turbulent and kinetic energy for each zone as a function of time. The largest 
values are attained for the local velocity minima (for a given mass zone), which is the dominant term in this quantity. The 
discrepancy is huge, the turbulent energy is 10-20 orders of magnitude less than the kinetic energy.  Even if the total turbulent energy could be converted to kinetic energy 
the modulation 
of the light curve would be still very small. However, the efficiency of the energy transfer is 
much lower than 100\%, therefore we conclude that -- at least in our
1D code -- the turbulent convection has a negligible effect in
altering the light curve shape from cycle to cycle, although it provides the necessary viscosity to control the amplitude. We note in passing that Buchler, Koll\'ath \& Cadmus (2004) found a much more significant turbulent energy in RV Tauri models. 

A logical step in this direction would be the use of multi-dimension hydrocodes, such as 
Mundprecht, Muthsam \& Kupka (2013), that would naturally enable the enhancement 
of turbulent energy or turbulent flux locally (see their Fig. 12) as opposed to the zone-averaged 
quantities in our 1D simulations. However, given the large difference we found between the kinetic 
and turbulent energies, it is highly questionable whether the required drastic change can occur in 
these overtone Cepheid models.

\subsection{Period change}

Several causes have been suggested for period variations in Cepheids,
both fundamental mode pulsators and overtone pulsators:

$\bullet$ {\bf Evolution} through the instability strip: This is 
almost certainly responsible for some of the period changes seen.  
While evolution does not necessarily proceed at a uniform pace 
(Fernie et~al. 1993), one direction of period change would
predominate and result in a parabolic $O-C$ diagram.

$\bullet$ {\bf Light-time effects} in binary systems: This produces 
cyclic apparent period changes. They must, however, be consistent with 
what is known about the orbit of the system. Possibly the best example 
is AW~Per (Welch \& Evans 1989). Light-time effect has been suggested
to explain the $O-C$ residuals in RT~Aur (Turner et~al. 2007).  
However, the velocity variations shown in Fig.~\ref{gamma} indicate no variation 
larger than $\pm$ 1 km\,s$^{-1}$ of the mean. This implies that an orbital 
velocity variation has not been observed  for data 
drawn from many different instruments over a long time interval.  

As discussed above, a substantial fraction (24\%) of overtone pulsators
have period variations characterized by alternate (though cyclic)
increases and decreases.  


Light-time effect is the {\it one} explanation for period change which 
is cyclic, but it is not viable for many of the overtone pulsators 
because of the scale and erratic nature. This implies that these 
quasi-cyclic period variations are caused by something in the pulsation 
process itself.  

$\bullet$ {\bf Star spots}: Neilson and Ignace (2014) have suggested
that the period variations of the Kepler Cepheid V1154 Cyg could be
produced by a hot spot on the surface caused by convection. In a study 
of yellow supergiants including several Cepheids,  Percy and Kim (2014)
suggest a similar possible cause for amplitude variation, large convection cells 
causing variation as the star rotates.    While starspots could affect
the time of maximum light, they would not have a cumulative effect as seen
in the $O-C$ diagram.

$\bullet$ {\bf Mass loss}: This has been suggested, for instance by
Neilson et~al. (2011, 2012), and worked out in detail. However, since
periods change {\it monotonically}, the quasi-cyclic variations
frequently seen in overtone pulsators are not due to mass loss. We
note, however, that the period changes in overtone pulsators discussed
in Sect.~5 which can reasonably be fit with parabolas all show
increasing periods. This is consistent with mass loss, although it 
may be only one of several factors.  

$\bullet$ {\bf Pulsation}:  Amplitude variation in 
Blazhko RR Lyr stars is not fully understood, but one possible explanation is
that it is  produced by 
high-order resonance (Buchler and Koll\'ath 2011).
Pulsation and excitation of a complicated group of 
modes may also play a role in Cepheid period change.  Percy and Kim (2014) 
suggest that convection may drive pulsation mode excitation and hence amplitude
variation, and the same  might also 
affect  Cepheid periods.  As discussed above, the fact that the level of pulsation 
instability on both short and long timescales appears to  depend on pulsation 
mode  suggests a role  for pulsation in period changes.

 The observed period changes   may be due to a combination
of these factors. However, the morphology of $O-C$ diagrams,
particularly for overtone pulsators, provides some clues. In
particular the high fraction of overtone pulsators which have
quasi-cyclic period variations is not consistent with stellar evolution
nor mass-loss. Furthermore, the size of the variations is too large to
uniquely originate from binary light-time effects. Hence the
pulsation process itself is indicated as a cause.

\section{Conclusions}

There are three primary results from the {\it MOST} observations of Cepheids.

$\bullet$  The observations of a fundamental mode pulsator (RT~Aur)
and an overtone pulsator (SZ~Tau) find greater instability in the
pulsation of the overtone Cepheid in the repetition of the light curve
and the Fourier parameters. 

$\bullet$ On the other hand, the deviations between cycles for both RT
Aur and SZ Tau follow a similar pattern as a function of phase of
increase after minimum radius and a return to a smaller value after
maximum radius.

$\bullet$ The $O-C$ curves indicate that on a time scale of
decades, the period changes of the overtone pulsator are more erratic.

Thus at both long and short time scales, the period variations of 
RT~Aur (fundamental mode) and SZ~Tau (overtone mode) differ, 
with the overtone mode   pulsator
exhibiting  greater instability at all time scales.

\section*{Acknowledgments} We are happy to thank Joseph E. Postma
for  his unpublished photometric data.  We also thank Zolt\'an
Koll\'ath for enlightening discussions.  Comments from an anonymous referee 
have improved the text, particularly in Section 4.   
This project has been supported by the `Lend\"ulet-2009 Young Researchers' 
Program of the Hungarian Academy of Sciences,  and the Hungarian OTKA
grant K83790. The research leading 
to these results has received funding from the European Community's 
Seventh Framework Programme (FP7/2007-2013) under grant agreement 
no. 269194 (IRSES/ASK)and No. 312844 (SPACEINN). Funding has also been 
received from the ESA PECS Contract No. 4000110889/14/NL/NDe.
RSz and AD were supported by the J\'anos Bolyai 
Research Scholarship of the Hungarian Academy of Sciences. RSz thanks 
the hospitality of CfA during a visit. Financial support for LSz was provided 
from the ESTEC Contract No.4000106398/12/NL/KML. Support for this work 
was also provided from the Chandra X-ray Center NASA Contract NAS8-03060 
(for NRE). JMM, DBG, AFJM and SR are grateful for financial aid from 
NSERC (Canada) and AFJM also to FRQNT (Quebec).  WWW was supported by the 
Austrian Science Fonds (FWF P22691-N16)
This research has made use of the SIMBAD database,
operated at CDS, Strasbourg, France.


\end{document}